\documentclass[a4paper,showpacs]{revtex4}
\usepackage{amsmath}
\usepackage{graphicx}
\begin{document}
\large
\title{
Nonlinear Schr\"odinger Equation with Spatio-Temporal Perturbations}

\author{Franz G. Mertens}
\affiliation
{Physikalisches Institut, Universit\"at Bayreuth, 95440 Bayreuth,
Germany}
\author{Niurka R. Quintero}
\affiliation
{Departamento de F\'isica Aplicada 1, E.U.P. Universidad de Sevilla, Virgen de \'Africa 7, 41011 Sevilla,
Spain}
\author{A. R. Bishop}
\affiliation
{Theoretical Division and Center for Nonlinear Studies, 
Los Alamos National Laboratory,  Los Alamos, NM 87545, USA}

\pacs{05.45.Yv, % Solitons
05.60.-k,  % Transport processes
 63.20.Pw % Localized modes}
}

\begin{abstract}
We investigate the dynamics of solitons of the cubic Nonlinear Schr\"odinger 
Equation (NLSE) with the following perturbations: non-parametric spatio-temporal driving of the form 
$f(x,t) = a \,\exp[i\, K(t)\, x]$, damping, and a linear term which serves to 
stabilize the driven soliton. Using the time evolution of norm, momentum and energy, or, alternatively, 
a Lagrangian approach, we develop a Collective-Coordinate-Theory which yields a set of ODEs for our 
four collective coordinates. 
These ODEs are solved analytically and numerically for the case of a constant, spatially periodic force $f(x)$. 
The soliton position exhibits oscillations around a mean trajectory with constant velocity. 
This means that the soliton performs, on the average, a unidirectional motion although the spatial 
average of the force vanishes. The amplitude of the oscillations is much smaller than the period of $f(x)$. 
In order to find out for which regions the above solutions are stable, we calculate the time evolution of the soliton momentum $P(t)$ and soliton velocity $V(t)$: This is a parameter representation of a curve $P(V)$ which is visited by the soliton while time evolves. Our conjecture is 
that the soliton becomes unstable, if this curve has a branch with negative slope. 
This conjecture is fully confirmed 
 by our simulations for the perturbed NLSE. 
Moreover, this curve also yields a good estimate for the soliton lifetime: the soliton lives longer, 
the shorter the branch with negative slope is.
\end{abstract}

%\begin{multicols}{2}
\maketitle
\section{Introduction} \label{uno}

The Nonlinear Schr\"odinger Equation (NLSE) is one of the paradigms of soliton physics, 
because it represents a completely
integrable system and has very many applications in practically all fields of physics, which are listed and discussed 
in several review articles \cite{kivmal, malo, fadd}. For applications it is important to study the perturbed 
NLSE 
\begin{equation} 
\label{aa}
iu_t + u_{xx} + 2|u|^2u = R[u (x, t); \, x, t]. 
\end{equation}
Many different kinds of perturbations $R$ have been considered and in 
particular the dynamics of a single soliton under these perturbations was investigated \cite{kivmal, malo}. 

In  this paper we consider  the following combination of perturbations
\begin{equation}
\label{ab}
R =f (x, t) - i \beta u (x, t) - \delta u (x, t)
\end{equation}
with the real parameters $\beta$  and $\delta $ and the {\it non-parametric spatio-temporal} driving force 
\begin{equation}
\label{ac}
f(x, t) = a e ^{iK(t)\cdot x }, 
\end{equation}
which yields several interesting effects, as we will see. The literature so far has mostly dealt 
with parametric driving 
\cite{kivmal,malo,bbk,bbb,scharf,igor1,potabk}. Non-parametric (external) driving was studied 
without space dependence, e. g., $f = \epsilon  \exp (i \omega  t)$ \cite{kaup,bs,bz}, and 
with a periodic space dependence, e.g., $f = \epsilon  \exp [i(kx - \omega t)]$ \cite{cohen,vyas}. 
Moreover, $f = \epsilon  \exp [i g(x,t)  - i \omega t]$, where $g$ is a function of $x - v t$, 
was considered, but no localized solutions were discussed 
\cite{vyas}. 

Our driving term in (\ref{ac}) was already used in the discrete form  $f_{n}(t) = a  \exp [i n \phi (t)]$. Here the integer 
$n$ denotes a lattice site 
in a nonlinear optical waveguide array which is modeled by a Discrete Nonlinear Schr\"odinger 
Equation (DNLSE) \cite{gorbach}. 
$\phi (t)$  is the incident angle of a laser beam. In order to obtain a ratchet effect, a biharmonic $\phi (t)$ was used which breaks a temporal symmetry. 

In this paper we work with an arbitrary function $K (t)$ in the driving term (\ref{ac}) and develop a 
Collective Coordinate (CC) Theory for the soliton dynamics which results in a set of nonlinear coupled ODEs for the CCs 
(Sections II and III).  
In order to obtain analytical and numerical solutions we then consider the case of temporally constant, 
spatially periodic driving $f (x) = a$ exp $(i Kx)$ with constant $K$ (Section IV). 
Although the spatial average of $f(x)$ vanishes, there is transport: the soliton performs  
a unidirectional motion on the 
average, in contrast to the case of the driving $R = V(x) u(x, t)$ 
with a periodic potential $V(x)$ in which the soliton performs an oscillatory motion around a minimum of $V(x)$ 
\cite{scharf}. Solutions of the CC-Eqs. for the case of a harmonic or biharmonic time dependence of $K (t) $ will be presented in a second paper.

The second term, $- i \beta u (x, t)$ with $\beta > 0$, in the perturbation (\ref{ab}) is a damping term which allows us 
to obtain a balance between the energy input from the driving and the dissipation.
Other more complicated damping terms have been considered in \cite{mal93,komi0203}.

The third term, $- \delta u(x, t)$, in Eq. (\ref{ab}) will turn out to be decisive for the stability of the driven soliton. For $\delta \ \ge 0$ the soliton radiates phonons (i.e., linear excitations) and eventually vanishes, 
or even breaks up into several solitons. For $ \delta < 0$ the situation is more complicated and will be discussed 
below. 

Section V presents a stable and an unstable stationary solution for the case without damping. 
For the region around the stable solution the CC-theory yields solutions in which all CCs exhibit oscillations with 
the same intrinsic frequency $\Omega$. 
However, tests by simulations, i.e. 
numerical solutions of the perturbed NLSE, reveal that the oscillatory solutions are stable only for certain regions of the 
initial conditions. These regions become broader when $\delta$ is more negative 
(Section VI). 

We conjecture that the stability of any of these oscillatory solutions can be predicted 
by our CC-theory by calculating the curve $P(V)$, where $P(t)$ and $V(t)$ are the momentum and 
the velocity of the soliton, respectively. This means that every point on this curve is visited 
during one period of the oscillatory solution. The conjecture is that the soliton will become 
unstable in a simulation, if the curve $P(V)$ has a branch with negative slope, and this is 
confirmed by our simulations (Section VI). Interestingly, the curve $P(V)$ not only 
predicts whether the soliton is 
unstable, but it also allows us to estimate the soliton lifetime: 
This time is longer, the shorter the branch with negative 
slope is (Section VI). 

The stability criteria for NLS-equations in the literature cannot be applied to our 
oscillatory solutions: The criterion of Vakhitov and Kolokolov \cite{vk,w} was established 
for stationary solutions, and the criterion of Barashenkov \cite{igor1,igor2} for solitons 
travelling with 
constant velocity. In \cite{igor1,igor2} the slope of a curve $\tilde{P}(\tilde{V})$, 
different from ours, is considered and 
 decides about the stability. However, here each soliton solution is represented 
by one point on this curve, i.e. the curve represents a family of solutions with different 
velocities. 

Finally we show in Section VII that the kinetic and canonical soliton momenta are identical, and we analytically 
calculate the soliton  and phonon dispersion curves.   

\section{Time evolution of norm, momentum, and energy} \label{dos}
Multiplication of the perturbed NLSE (\ref{aa}) by $u^*$ and subtraction of the complex conjugate NLSE multiplied by $u$ yields
\begin{equation}
\label{ad}
\frac{\partial \varrho}{\partial t} +  \frac{\partial j}{\partial x} = i (R^* u - R u^*)
\end{equation}
with the density and current density
\begin{equation}
\label{ae}
\varrho = |u|^2 \, , \, j = i (u^*_x u - u^* u_x).
\end{equation}

By integration of Eq.(\ref{ad}) over $x$, assuming decaying boundary conditions, we obtain the time evolution of the norm,
\begin{equation}
\label{af}
N = \int ^{+\infty} _{-\infty} dx |u|^2 \, ,
\end{equation}
which is
\begin{eqnarray}
%\label{ag}
\nonumber 
\dot{N} &=& i \int ^{+\infty} _{-\infty} dx (R^* u - R u^*) \\
\label{aag}
\,\ & =& - 2 \beta N + i \int ^{+\infty} _{-\infty} dx (f^* u - f u^*) \, ,
\end{eqnarray}
where the dot denotes the time-derivative and the terms with $\delta$ have dropped out.

Multiplication of Eq. (\ref{aa}) by $u_x^*$, addition of the complex conjugate equation  multiplied by $u_x$, and integration over $x$ yields the time evolution of the momentum
\begin{equation}
\label{ai}
\dot{P} = \int ^{+\infty} _{-\infty} dx (R^* u_x + R u_x^*)
\end{equation}
where $P$ is defined as 
\begin{equation}
\label{aj}
P = \frac{i}{2 }\int ^{+\infty} _{-\infty} dx (u u^*_x -  u^* u_x) \, .
\end{equation}

Multiplication of Eq. (\ref{aa}) by $u^*_t$, addition of the complex conjugate equation multiplied by $u_t$, and integration yield the time evolution of the energy
\begin{equation}
\label{ak}
\dot{E} = - \int ^{+\infty} _{-\infty} dx (R^* u_t + R u_t^*) \, ,
\end{equation}
where
\begin{equation}
\label{al}
E = \int ^{+\infty} _{-\infty} dx [|u_x|^2 - |u|^4] \, .
\end{equation}

Interestingly, for $\beta = 0$ and time independent force, 
i. e. $R = f(x) - \delta u$, the r.h.s. of Eq. (\ref{ak}) can be written as a time derivative 
\begin{equation}
\label{am}
\dot{E} = - \frac{\partial}{\partial t} \int ^{+\infty} _{-\infty} dx [f^* u + f u^* - \delta |u|^2] \,.
\end{equation}
Thus the {\em perturbed} NLSE (\ref{aa}) without damping possesses a conserved quantity for {\em arbitrary} $f(x)$
\begin{equation}
\label{an}
E^{tot}  =  \int ^{+\infty} _{-\infty} dx [|u_x|^2 - |u|^4 - \delta |u|^2 + f^* u + f u^*] \,.
\end{equation}

$f(x)$ can be interpreted as a constant (external) force, in contrast to the case of the driving 
$R = V(x) u$, where $V (x)$ can be understood as a potential (in which the solitons move). In this case the conserved quantity is \cite{scharf} (see also \cite{prlbishop})
\begin{equation}
\label{ao}
E_{para}^{tot} =  \int ^{+\infty} _{-\infty} dx [|u_x|^2 - |u|^4 + V(x) |u|^2 ] \,.
\end{equation}

In Sections \ref{cuatro} and \ref{cinco} 
we will show that a soliton under a constant {\em periodic} force $f(x) = a \exp (iKx)$ performs, on the average, a 
{\em unidirectional} motion, although the average of $f$ vanishes.

When we include the damping $(\beta > 0)$, the time evolution of the total energy is 
\begin{equation}
\label{ap}
\dot{E}^{tot}  = - \beta \int ^{+\infty} _{-\infty} dx [2|u_x|^2 - 4 |u|^4 - 2 \delta |u|^2 + f^* u + f u^*] \,.
\end{equation}

\section{Collective Coordinate Theory} \label{tres}

The 1-soliton solution of the unperturbed NLSE reads \cite{kivmal}
\begin{equation}
\label{aq}
u(x, t) = 2 i \eta \, \mathrm{sech} [2 \eta (x- \zeta)] e^{-i(2 \xi x + \phi)}
\end{equation}
with the real parameters $\eta > 0$ and $\xi $, the soliton position
\begin{equation}
\label{ar}
\zeta (t) = \zeta _0 - 4 \xi t\, ,
\end{equation}
and the phase of the internal oscillation
\begin{equation}
\label{as}
\phi (t) = \phi _0 + 4 (\xi ^2 - \eta ^2) t \, .
\end{equation}
The soliton has amplitude $2 \eta$, width $1/(2 \eta)$, velocity $V = - 4 \xi $ and phase velocity of 
the carrier wave $V_{ph} = - 2 \xi + 2 \eta ^2/\xi $ .

Including the term with $\delta $ on the r.h.s. of Eq. (\ref{aa}), only the phase is changed: 
\begin{equation}
\label{at}
\phi (t) = \phi _0 + [4 (\xi ^2 - \eta ^2) - \delta ] t \, .
\end{equation}

For this reason the term with $\delta $ need not be treated as a perturbation and will be counted among the unperturbed parts of the NLSE in the following
\begin{equation}
\label{au}
i u_t + u_{xx} + 2 |u|^2 + \delta u = f (x, t) - i \beta u \, .
\end{equation}
Therefore we now define the energy as 
\begin{equation}
\label{av}
E =  \int ^{+\infty} _{-\infty} dx [|u_x|^2 - |u|^4 - \delta  |u|^2 ] \,,
\end{equation}
which is conserved for the unperturbed NLSE. Using Eq. (\ref{aq}) we obtain the soliton energy
\begin{equation}
\label{aw}
E^{sol} = 16 \eta \xi ^2 - \frac{16}{3} \, \eta ^3 - 4 \delta \eta \, .
\end{equation}

The definitions (\ref{af}) and (\ref{aj}) need  not be changed, because all terms with $\delta $ drop out on the r.h.s. of Eq. (\ref{aag}) and (\ref{ai}). The norm and the momentum of  the soliton are, respectively, 
\begin{equation}
\label{ax}
N = 4 \eta\, ; \, P =- 8 \eta \xi \, .
\end{equation}
Writing $P = M V$, the soliton mass  is related to the norm by $M = N/2$.

From the Inverse Scattering Theory (IST) it is well known \cite{kivmal} that a perturbation theory results in a time dependence of $\eta$ and $\xi $ and in a change of the simple time dependence of $\zeta $ and $\phi $ in Eqs. (\ref{ar}) and  (\ref{as}). Therefore we take Eq. (\ref{aq}) as an ansatz with the four Collective Coordinates (CCs) $\eta (t), \xi (t), \zeta (t)$ and 
$\phi (t)$ (a similar ansatz has been used for optical solitons \cite{chaos,komi04})  
and insert it into Eqs. (\ref{aag}), (\ref{ai}) and (\ref{ak}), 
using $f (x, t) = a \exp [i K (t) \cdot x]$ from Eq. (\ref{ac}).

Eq. (\ref{aag}) yields
\begin{equation}
\label{ay}
\dot{\eta} = - 2 \beta \eta - a  \frac{\pi}{2}  \mathrm{sech} A \cdot \cos B, 
\end{equation}
with
\begin{eqnarray}
\label{az}
A(t) &=& \frac{\pi}{4} [K(t) + 2 \xi (t)] / \eta (t), \\
\label{ba}
B(t) &=& \phi (t) + [K(t) + 2 \xi (t)] \zeta (t) \, .
\end{eqnarray}

Two of the three terms which result on the r.h.s. of (\ref{ai}) cancel with the term $-8\dot{\eta} \xi $ on the l.h.s.. The remaining terms are
\begin{equation}
\label{bb}
\dot{\xi} = a A \, \mathrm{sech}
 A \cdot \cos B \, .
\end{equation}

Finally, Eq. (\ref{ak}) gives $8 \eta \dot{\xi} \{...\} - 4 \dot{\eta} \{...\} = 0$ with two 
different curly brackets. This equation can be fulfilled by 
setting both curly brackets to zero, which yields 
\begin{eqnarray}
\label{bc}
\dot{\zeta}& =& - 4 \xi + \frac{a \pi ^2}{8 \eta ^2} \mathrm{sech} A \cdot \tanh A \cdot \sin B, \\
\label{bd}
\dot{\phi} + 2 \zeta \dot{\xi}& = & 4 (\xi ^2 - \eta ^2) - \delta + \frac{a \pi A}{2 \eta} \mathrm{sech} 
A \cdot \tanh A \cdot \sin B\,.
\end{eqnarray}

We can alternatively use the Lagrangian method which will also be useful for section \ref{siete}. 
Our perturbed NLSE is equivalent to an Euler-Lagrange-Equation, generalized by a dissipative term, 
\begin{equation}
\label{be}
\frac{d}{dt}\, \frac{\partial \mathcal{L}}{\partial u_t^*} + \frac{d}{dx} \, \frac{\partial \mathcal{L}}{\partial u_x^*} - \frac{\partial \mathcal{L}}{\partial u^*} = \frac{\partial \mathcal{F}}{\partial u_t^*}, 
\end{equation}
with the Lagrangian density
\begin{equation}
\label{bf}
\mathcal{L}= \frac{i}{2} (u_t u^* - u_t^* u) - |u_x|^2 + |u|^4 + \delta |u|^2 - f u^* - f^*u, 
\end{equation}
and the dissipation function
\begin{equation}
\label{bg}
\mathcal{F} = i \beta (u u_t^* - u^* u_t) \, .
\end{equation}

Inserting our CC-ansatz and integrating over the system we obtain the CC-Lagrangian
\begin{equation}
\label{bh}
L = 4 \eta \dot{\phi} + 8 \eta \zeta \dot{\xi} - 16 \eta \xi ^2 + \frac{16}{3} \, \eta ^3 + 4 \delta \eta - 2\pi a \, \mathrm{sech} A \cdot \sin B
\end{equation}
and the CC-dissipation function
\begin{equation}
\label{bi}
F = - \beta (8 \eta \dot{\phi} + 16 \eta \zeta \dot{\xi}) \, .
\end{equation}

The CC-equations are then obtained by the four generalized Lagrange equations
\begin{equation}
\label{bj}
\frac{d}{dt} \, \frac{\partial L}{\partial \dot{\psi}} - \frac{\partial L}{\partial \psi} = 
\frac{\partial F}{\partial \dot{\psi}}, 
\end{equation}
where $\psi$ stands for the four CCs $\eta, \xi, \zeta$ and $\phi $. The resulting four ODEs are identical with the Eqs. (\ref{ay}), (\ref{bb}), (\ref{bc}) and (\ref{bd}).

Finally we evaluate Eq. (\ref{an}), yielding  
\begin{equation}
\label{bk}
E^{tot} = 16\eta \xi ^2 - \frac{16}{3}  \eta ^3 - 4 \delta \eta + 2 \pi a \,  \mathrm{sech}  A \cdot \sin B\, .
\end{equation}
Here the first three terms are the soliton energy (\ref{aw}), while the last term stems from the perturbations. We note that $E^{tot}$ is conserved only in the case of no damping $(\beta = 0)$ and time independent force $f(x) = a \exp (iKx)$ with constant $K$, see below Eq.  (\ref{am}).

\section{Constant, spatially periodic force} \label{cuatro}

We consider this case because it exhibits some surprising, counter-intuitive features: E. g., the soliton position 
performs oscillations on a length scale that is very different from the spatial period $L$ of the force.

We take a constant $K$ in Eq. (\ref{ac}), i. e. $f(x) = a \exp (iKx)$, and consider only small values of $|K|$ such 
that the period $L = 2 \pi/|K|$ is much larger than the soliton width. We first consider the case with damping 
($\beta > 0$) for which we can expect  ``steady-state 
solutions'' (see below Eq.\ (\ref{bo})) for times much larger than a transient time $ \tau $ on the order of $1/\beta $. 

The transformation $u(x,t)=\Psi(X,t)\,\exp(i K x)$  into a moving frame $X = x - V_{f} t$ with $V_{f} = 2 \,K$ leads 
to the autonomous equation 
\begin{equation}
\label{auto}
i \Psi_t + \Psi_{XX} + 2 |\Psi|^2 \Psi=  a + (K^2 - \delta) \Psi  - i \beta \Psi,
\end{equation}
with the non-parametric constant driving term $a$.  
Apart from the factor $K^2 - \delta := c^{2}$, which can be eliminated by scaling time by $c^{2}$, space 
by $c>0$, and $\Psi$ by $1/c$, Eq.\ (\ref{auto}) is the same as an autonomous equation, which was obtained from 
the NLSE (\ref{aa}) with $R=\epsilon \exp(i \omega t) - i \beta u$ by the substitution 
$u=\Psi(x,t) \exp(i \omega t)$, setting $\omega=1$ \cite{bs,bz}. 
However, these investigations differ from ours in several respects: In Ref.   \cite{bs} static soliton solutions of Eq.\ 
(\ref{auto}) were obtained numerically, then an existence and stability chart was constructed on the 
$(a,\beta)$ plane. In Ref. \cite{bz} a singular perturbation expansion was performed at the soliton's 
existence threshold. In contrast, we study moving solitons by solving our CC-equations, and test the results by simulations, i.e. by numerically solving the NLSE (\ref{aa}). Besides the 
steady-state solutions for $\beta>0$ (this section), for the case $\beta=0$ we study stationary solutions and oscillatory 
solutions, which 
have a more complicated time dependence (Section V). 

The CC-equations from the previous section yield steady-state solutions,  
in which the driving is compensated by the damping, using the ansatz 
$\zeta = \zeta _0 + V_f t$ with $V_{f}=2 K$ 
and constant $\eta$, $\xi$, and $\phi$: 
\begin{eqnarray}
\label{bm}
u_{f}^{\pm}(x,t) = 2 i \eta_{f} \mathrm{sech}\{2 \eta_{f} [x - \zeta(t)] \} 
\mathrm{e}^{i (K x- \phi_{f}^{\pm})}, 
\end{eqnarray}
with 
\begin{eqnarray}
\label{bn}
\eta_f &=& \sqrt{K^2 - \delta } /2, \\
\label{bo}
\phi _{f}^{\pm} &=& \pm \frac{\pi}{2} + \, 
\arcsin \, \frac{4 \beta \eta_f}{\pi a}.                                       \end{eqnarray}
We denote this as steady-state solutions, because $\phi$ is 
constant, in contrast to stationary solutions, where $\phi$ has a linear time dependence 
(Section V). 
These solitons have an internal structure due to the factor $\exp(i K x)$, but no internal oscillations since $\phi_{f}^{\pm}$ is constant. 
In the moving frame these solitons correspond to the above mentioned static 
solutions of Eq.\ (\ref{auto}) in Ref. \cite{bs}. 

We have numerically solved the CC-equations for many sets of initial conditions  (IC)  
  $\eta_0, \zeta_0, \xi_0$ and $\phi_0$. 
There is a basin of attraction around the solution (\ref{bm}) with 
$\phi_{f}^{+}$. The solitons always evolve to this solution, 
except when the  values of $\eta _0, \xi _0$ and $\phi _0$ are 
too far from those of this steady-state solution and when the damping $\beta$ is too large. An example 
for this is the parameter set $a=0.05$, $K = 0.01$, $\delta = -3$, $\beta \ge \beta_{c} \approx 0.003$ with the IC 
 $\eta_0 = 1$,  $\xi_0 = \zeta_0 = \phi_0 = 0$. Here the soliton vanishes, i.e. its amplitude and energy go to 
zero while its width goes to infinity. In order to achieve a convergence to 
the stable steady-state solution 
one can either reduce 
$\beta$ below the critical value $\beta_{c}$ (which depends on the other parameters and the IC), or go closer 
to $\eta_{f}=0.8660$ by reducing $\eta_{0}$ by $0.1$, for instance, or go closer to $\phi_{f}^{+}$ by choosing 
$\phi_{0}=\pi/2$.

In this context it is interesting to consider the total energy (\ref{bk}) and its time derivative in which the CC-equations can be inserted. Using MATHEMATICA \cite{mathe} we obtain
\begin{equation}
\label{bp}
\dot{E}^{tot} (t) = - \beta \{8\eta [4( \xi ^2 - \eta ^2) -  \delta ] + 4 \pi  A \,  \mathrm{sech}  A \cdot \tanh A \cdot \sin B \} \, .
\end{equation}
This is indeed zero for the steady-state solution because $\xi_{f} = - K/2$ and thus $A \equiv 0$. On the path 
to that solution, Eq.\ (\ref{bp}) alternately exhibits both signs. I. e., the 
total energy performs oscillations which become smaller and smaller while approaching the final value $E^{tot}_f$.

According to Eq. (\ref{bn}), $\delta < K^2$. As we choose $|K| \ll 1$ (see above), $\delta$ is either positive but 
very small, or $\delta$ is negative. This results from the CC-theory; in section \ref{siete} 
we will show that the driven 
soliton can be stable only for $\delta < 0$, by taking into account the phonon modes.

\begin{figure}[ht!]
\begin{center}
\begin{tabular}{cc}
\ & \\
\includegraphics[width=8.0cm]{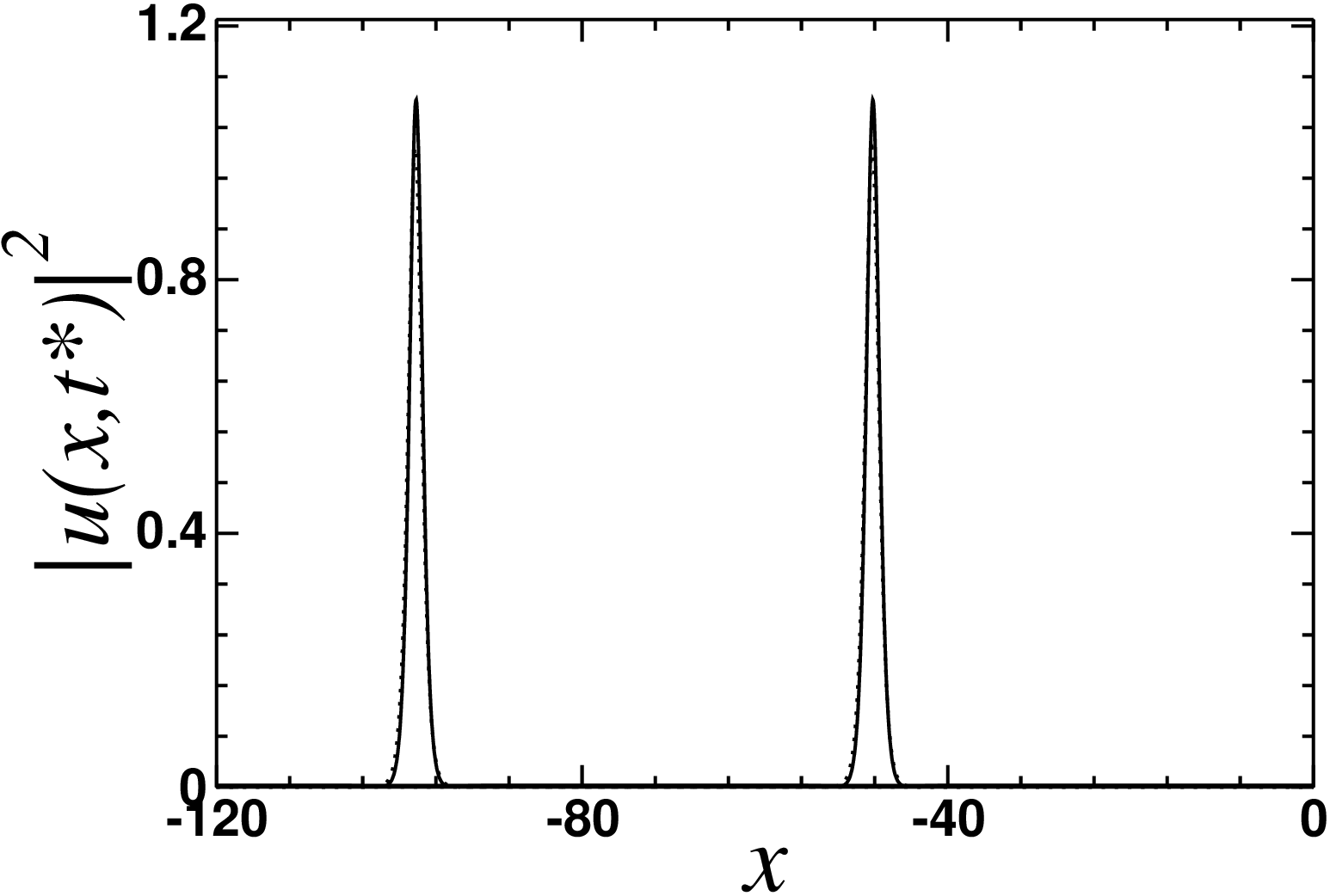}  & 
\quad \includegraphics[width=8.0cm]{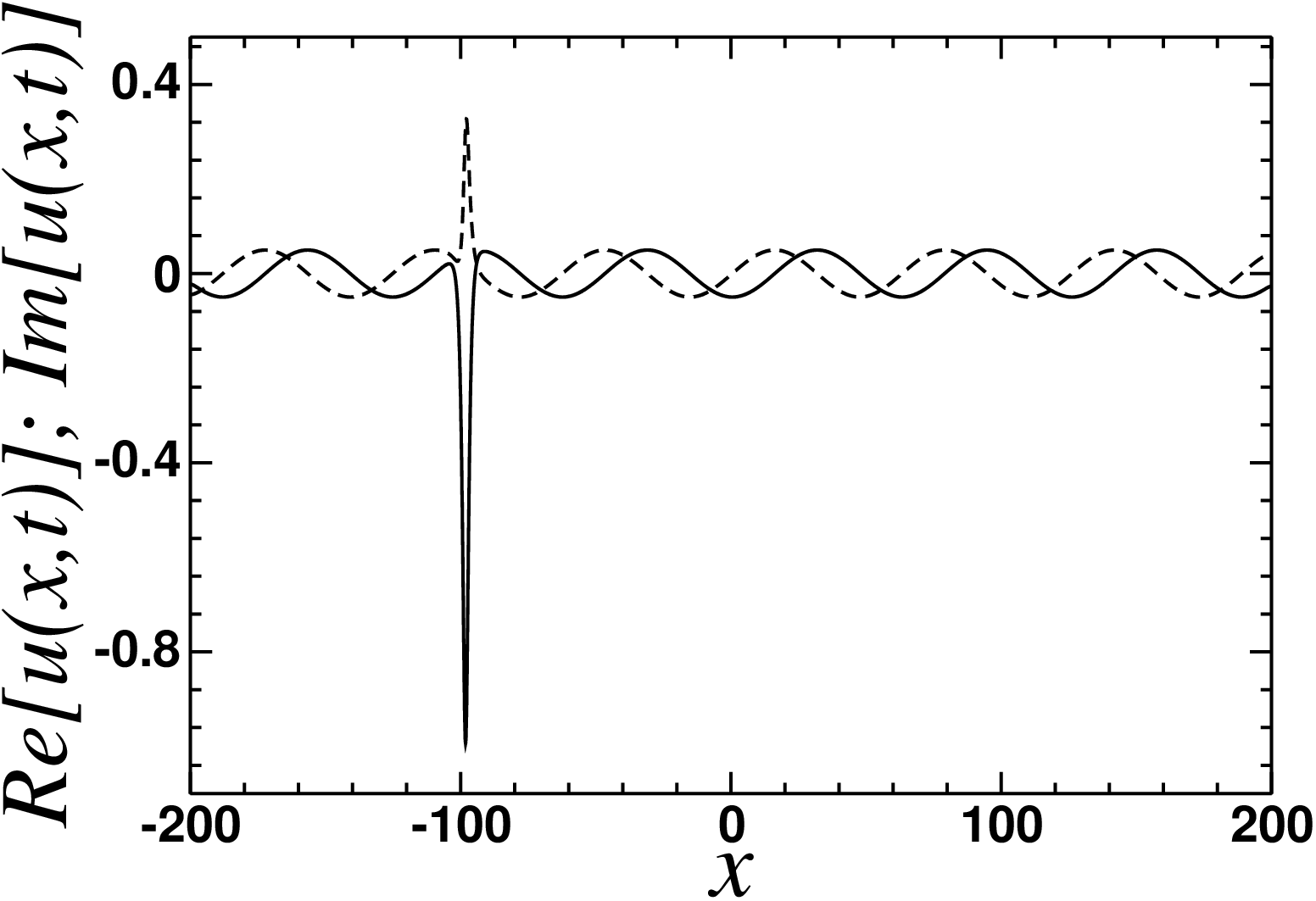}
\end{tabular}
\end{center}
\caption{Left panel: Soliton moving to the left for $t^{*}=250, 500$. Simulations of NLSE (solid lines) and numerical 
solutions of CC-equations (dotted lines). 
Right panel: Real (solid line) and imaginary (dashed line) parts of $u(x,t)$ for $t=500$. 
Parameters:  
$K=-0.1$, $a=0.05$, $\delta=-1$, $\beta=0.05$, with IC $\xi_{0}=0$, $\zeta_{0}=0$,    
$\phi_{0}=1.69$ and $\eta_{0}=0.5$. 
}
\label{fig1a} 
\end{figure}

\begin{figure}[ht!]
\begin{center}
\begin{tabular}{cc}\ & \\
\includegraphics[width=8.0cm]{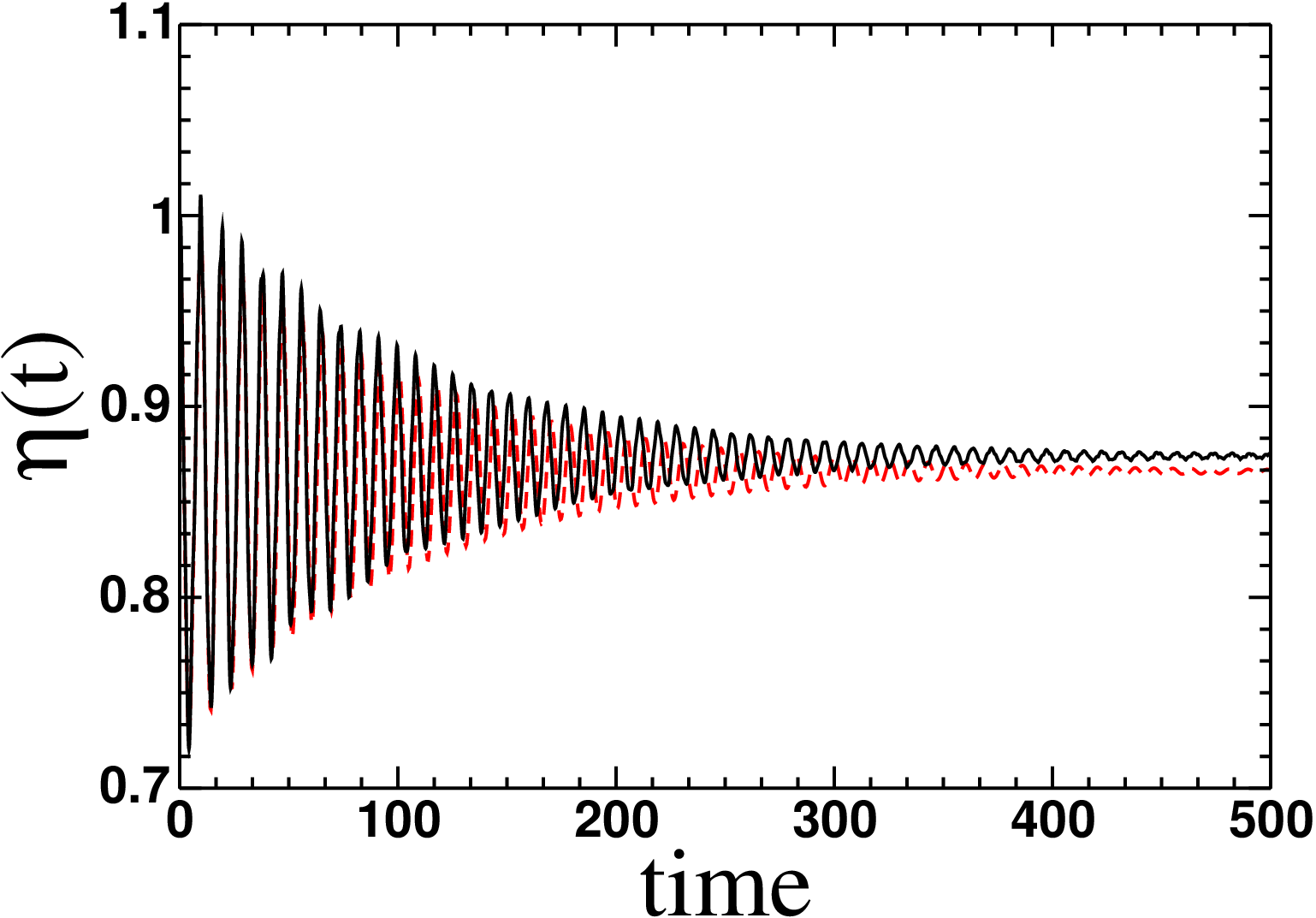}  & 
\quad \includegraphics[width=8.0cm]{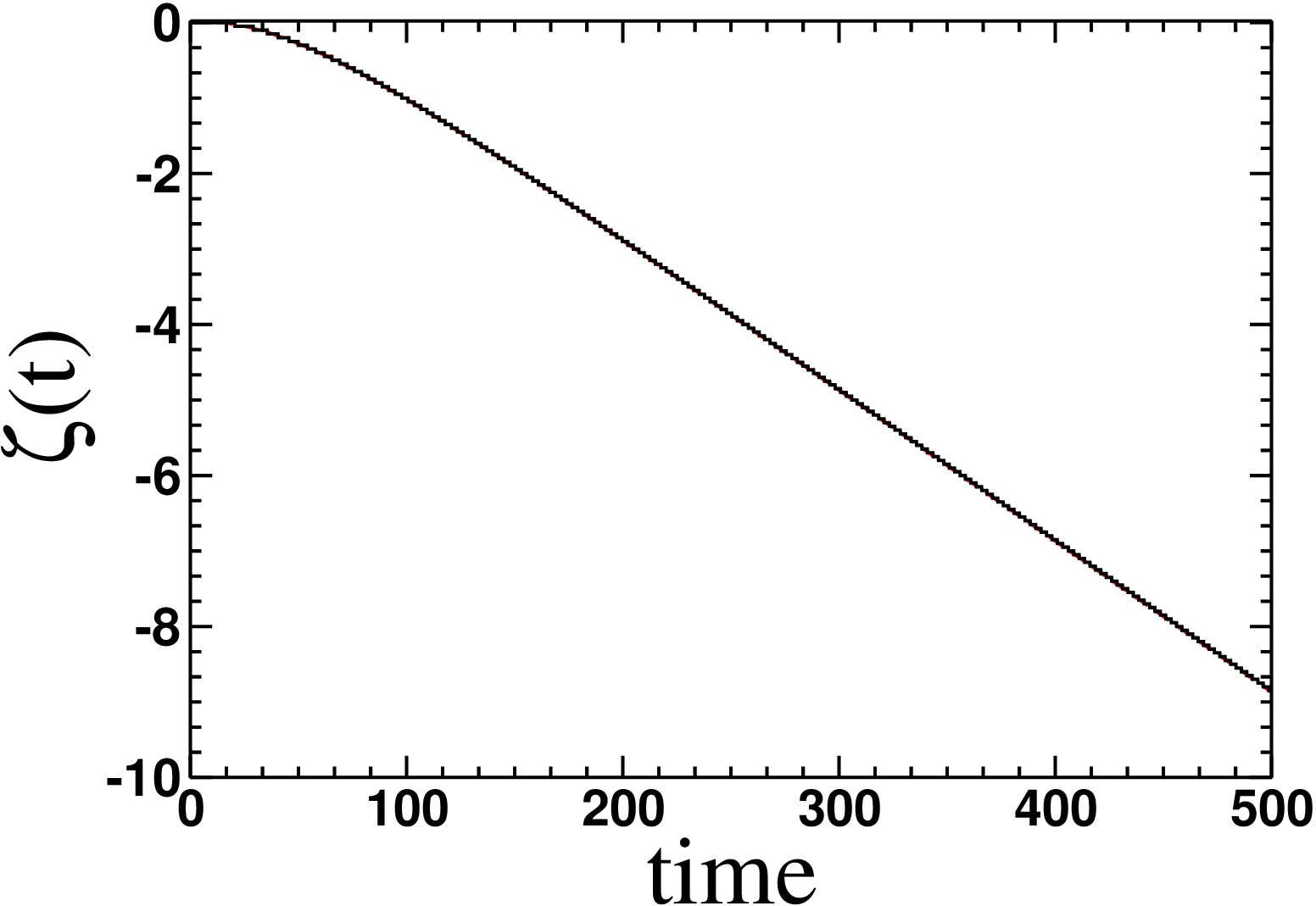}
\end{tabular}
\end{center}
\caption{(Color online). The amplitude and position of the soliton 
obtained from a simulation of the NLSE (red dashed lines) 
and from the numerical solution 
of the CC equations (solid lines). 
Parameters: $K=-0.01$, $a=0.05$, $\delta=-3$, $\beta=0.01$, with IC $\xi_{0}=0$, $\zeta_{0}=0$,    
$\phi_{0}=\pi/2$ and $\eta_{0}=1$. 
}
\label{fig1} 
\end{figure}

The quality of the CC-theory must be tested by simulations. 
The soliton shape agrees very well (Fig. \ref{fig1a}), but in the simulations the soliton resides on a small constant 
background because the perturbation $f(x) = a \exp(i K x)$ does not vanish far away from the soliton. This background 
is 
\begin{equation} \label{backg}
u_{bg} = - \frac{a}{\omega_{K}} \, \mathrm{e}^{i K x}, 
\end{equation} 
with $\omega_{K}= K^2 -\delta -i \beta$; see also the last term in Eq.\ (\ref{nlse2}). Plotting the real and 
imaginary parts of $u$, the spatial period $L= 2\pi/|K|$ is observed. The norm density 
$|u|^2$ forms a shelf on which the soliton moves; the shelf height is quantitatively confirmed. 
The dynamics of the soliton is practically not affected by 
the background: the time evolution of the soliton position is identical in the CC-theory and the simulations 
(Fig.\ \ref{fig1}, right panel); only the soliton amplitude differs a little (left panel).

\begin{figure}[ht!]
\begin{center}
\begin{tabular}{cc}\ & \\
\includegraphics[width=7.5cm]{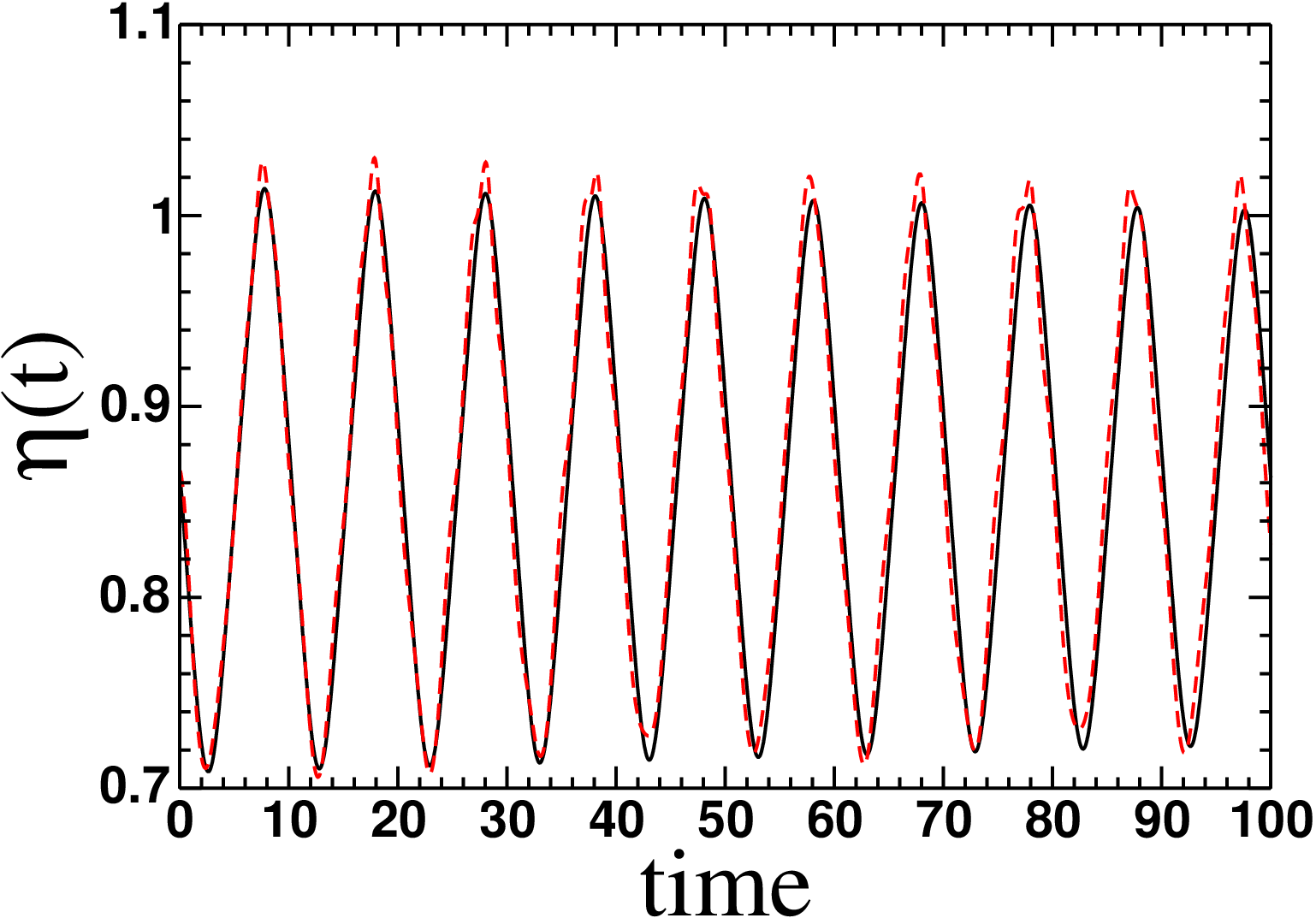} & 
\quad \includegraphics[width=7.5cm]{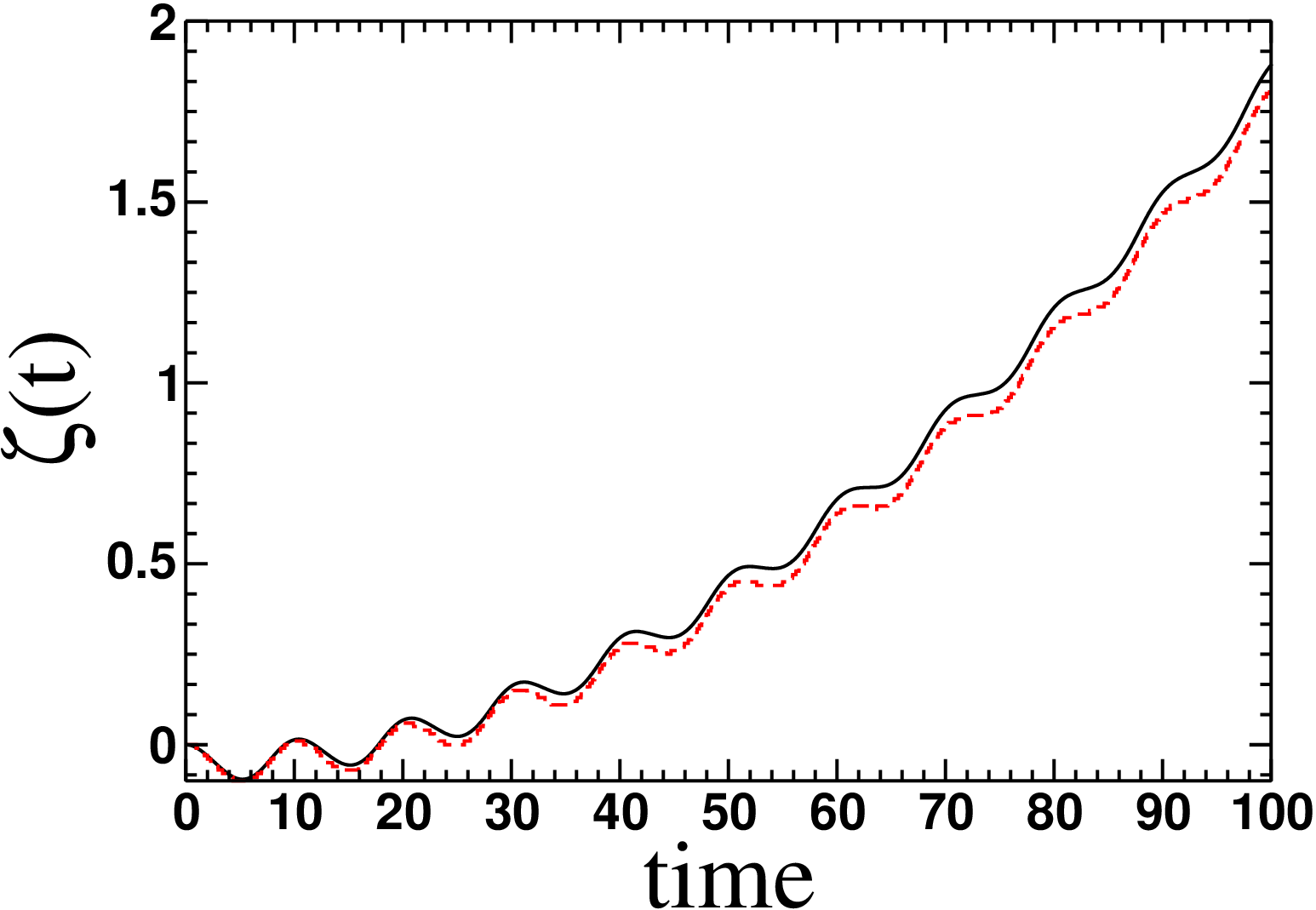}
\end{tabular}
\end{center}
\caption{
(Color online).  The amplitude and position of the soliton 
obtained from a simulation of the NLSE (red dashed lines) 
and from the numerical solution of the CC equations (solid lines). 
Parameters: $K=0.1$, $a=0.05$, $\delta=-3$, $\beta=0.001$, with IC $\xi_{0}=0$, $\zeta_{0}=0$, 
$\phi_{0}=0$ and $\eta_{0}=\sqrt{K^{2}-\delta}/2$.  
} 
\label{fig2}
\end{figure}

All CCs exhibit decreasing oscillations with an intrinsic frequency $\Omega $ which will be discussed in the 
next section. However, oscillations are not visible in the soliton position $\zeta (t)$ in Fig. \ref{fig1}  because 
the linear term dominates the time evolution. By reducing the damping $\beta $ and by choosing 
a smaller time scale one can see the oscillations also in $\zeta (t)$, see Fig. \ref{fig2}.

In any case, the soliton performs, on the average, 
a unidirectional motion although the spatial average of the periodic force 
$f (x) = a \, \exp (iKx)$ is zero. Thus, this is a ratchet-like system in which the translational symmetry is broken 
by the inhomogeneity $f(x)$ in the NLS-equation.  

The period $L = 2 \pi/|K| $ of the inhomogeneity $f(x)$ is not reflected in the soliton 
dynamics, because our Eqs.\ (\ref{aa})-(\ref{ac}) could be reduced to the autonomous Eq.\ 
(\ref{auto}). For early times $(t \le \tau = O(1/\beta ))$ the soliton performs the above mentioned oscillations. However, the amplitude of these oscillations is much smaller 
than $L$ (e.g., $L = 62.8$ for $K = 0.1)$. When the strength $a$ of the inhomogeneity is strongly increased $(a = 1$, for instance), $\zeta (t)$ exhibits a staircase structure. The height of the steps is larger than the amplitude of the above oscillations, but still much smaller than $L$. However, such values of $a$ represent a strong perturbation for which the CC-theory can no longer be valid. Indeed, in the simulations the soliton soon becomes unstable.

\section{Oscillatory solutions} \label{cinco}

In order to study in more detail the intrinsic oscillations in the CCs (see previous section), we consider the case without damping $(\beta = 0)$. Here oscillatory 
solutions are possible because the total energy (\ref{bk}) is conserved, see Eq. (\ref{bp}). This means that in Eq. (\ref{bk}) oscillations of the soliton energy, Eq. (\ref{aw}), are compensated by the oscillations of the term $2 \pi a \, \mathrm{sech} A (t) \cdot \sin B (t)$ stemming from the perturbations. This is confirmed by inserting numerical solutions of the CC-equations 
into Eq.\ (\ref{bk}).

In order to obtain analytical solutions our approach is to look first for stationary solutions and then to consider small oscillations around them. For the stationary solutions we make the ansatz $\zeta = \zeta _s + V_s t, \eta = \eta_s, \xi = \xi _s$ and $\phi = \phi _s - \alpha _s t$. The CC-equations (\ref{ay}), (\ref{bb}), (\ref{bc}), (\ref{bd}) yield

\begin{eqnarray}
\label{bq}
0 &=& -  \frac{a \pi }{2} \, \mathrm{sech} A_s \cdot \cos B, \\
\label{br}
0 &=&  a A_s \, \mathrm{sech} A_s \cdot \cos B, \\
\label{bs}
V_s &=& - 4 \xi _s + \frac{a \pi ^2}{8 \eta^2_s} \, \mathrm{sech} A_s \cdot \tanh A_s \cdot \sin B, \\
\label{bt}
- \alpha _s &=& 4  (\xi _s^2 - \eta ^2_s) - \delta  + \frac{a \pi A_s}{2 \eta_s} \, 
\mathrm{sech} A_s \cdot \tanh A_s \cdot \sin B,
\end{eqnarray}
with
\begin{eqnarray}
\label{bu}
A_s &=& \frac{\pi}{4} (K + 2 \xi _s)/\eta _s, \\
\label{bv}
B (t) &=& \phi _s + (K + 2 \xi _s)\zeta _s + [(K + 2 \xi _s) V_s - \alpha _s] t \,.
\end{eqnarray}

Eq. (\ref{bq}) must be fulfilled for arbitrary $t$, thus $\cos B \equiv 0$, which leads to
\begin{eqnarray}
\label{bw}
\alpha _s &=& (K + 2 \xi _s) V_s, \\
\label{bx}
\phi _s + (K + 2 \xi _s)\zeta _s &=& \pm \frac{\pi}{2}, \\
\label{by}
\sin B &=& \pm 1, 
\end{eqnarray}
respectively. 
We distinguish two different cases: In case I, $\alpha_{s}=0$ and we obtain the steady-state 
solutions of Section IV. In the co-moving frame these solutions 
correspond to 
two exact static solutions of (\ref{auto}) for zero damping 
\cite{bs,igor3,igor4}. Here one solution is  stable below a
critical driving strength, whereas the other one is always 
unstable. 

In case II, $\alpha_{s} \ne 0$, which means that we obtain stationary solutions. Here 
 we can restrict ourselves to the case $\zeta _s = \xi _s = 0$, because other values yield 
qualitatively similar results. Using Eqs. (\ref{bs}), (\ref{bt}), (\ref{bu}) and (\ref{bw}) 
one transcendental equation for $\eta _s$ remains:  
\begin{equation}
\label{eq-eta}
4 \eta_{s}^{2} = -\delta \pm \frac{\pi^{2} a K}{4 \eta_{s}^{2}}  \mathrm{sech} A_s \cdot \tanh A_s.   
\end{equation}
For the parameter set $a=0.05$, $K=0.1$ and $\delta=-3$ we get $\eta_{s}=0.866239$ for $\phi_{s}=\pi/2$ and 
$0.865811$ for $\phi_{s}=-\pi/2$. The numerical solution of the CC-equations for these two cases reveals 
that in the former case we have a stable stationary solution, whereas in the latter case 
the solution is unstable.  

For the region around the stable stationary solution we expect that all CCs exhibit oscillations 
and we assume that these oscillations are harmonic if the amplitudes are 
sufficiently small. We choose $\phi_{0}=\pi/2$ and make the ansatz
\begin{eqnarray} \label{eq1-ansatz}
\zeta(t)&=& \bar{V} t - a_{\zeta}\, \sin(\Omega t), \\   \label{eq2-ansatz}
\eta(t)&=& \eta_{0} + a_{\eta}\, [1- \cos(\Omega t)], \\  \label{eq3-ansatz}
\xi(t)&=&  - a_{\xi}\,[1- \cos(\Omega t)], \\ \label{eq4-ansatz}
\phi(t)&=&\phi_{0} - \alpha t+ a_{\phi} \sin(\Omega t). 
\end{eqnarray}
This ansatz takes into account that the soliton first starts its oscillatory motion (i.e., $\zeta-\bar{V} t$ is linear 
in $t$ for small $t$), and then the soliton shape changes (i.e., $\eta$ is quadratic for small $t$).

Since we have considered the 
stationary solution that belongs to $\xi_{s}=\zeta_{s}=0$, we can neglect $\xi$ compared to $K$ and obtain 
for $B$ in Eq.\ (\ref{ba}): 
\begin{equation}
\label{sol-B}
B= \phi_{0} + (a_{\phi} - K a_{\zeta}) \sin(\Omega t)+(K \bar{v} -\alpha) t.  
\end{equation}
$K a_{\zeta}$ is one order smaller than $a_{\phi}$, because $|K| \ll 1$, and we get 
\begin{equation}
\label{sol1-B}
B= \phi_{0} + a_{\phi} \, \sin(\Omega t)+(K \bar{V} -\alpha) t.  
\end{equation}
Here we can distinguish two limiting cases: In case 1, the linear terms cancel and we have a pure oscillatory behavior. 
In case 2,  the linear term dominates the oscillatory term which can then be neglected.  
Indeed, taking the second derivative of $\phi$ with respect to time and using  (\ref{ay}), 
(\ref{bb}), (\ref{bc}) and (\ref{bd}) one obtains 
$\ddot{\phi}=c_{1} \sin(B) + c_{2} \sin(2\,B) + c_{3} \cos(B)+c_{4} \cos(2\, B)+c_{5}$, where 
$c_{i}$ with $i=1,2,...,5$ are functions of the CCs. Considering the above approximations, and in addition 
when the terms of order of $a K \bar{V} (4 \eta_{0}^2+\delta)\,t$ are negligible, 
one realizes that $c_{1}$, $c_{2}$, $c_{4}$ and $c_{5}$ are small compared with 
$c_{3}=\Omega^2$, 
\begin{equation} \label{eq-ome}
\Omega^{2} \approx 4 \pi a \eta_{0},  
\end{equation} 
and hence, the Eq.\ for $\phi(t)$ reads 
\begin{equation} \label{eq-phi1}
\ddot{\phi} - \Omega^{2} \cos(\phi)=0. 
\end{equation} 
Integrating this equation once we obtain
\begin{equation} \label{eq-phi2}
\dot{\phi} = \pm \Omega \sqrt{2}\, \sqrt{\sin(\phi)+C},
\end{equation}  
where $C=\dot{\phi}^{2}_{0}/(2\,\Omega^{2})-\sin(\phi_{0}) \approx ((4 \eta_{0}^{2}+\delta)^2)
/(8 \pi a \eta_{0})-1$. Then, if $C \gg 1$,  the constant term dominates and so  
$\phi(t)$ goes linearly, i.e. $\phi(t)=\pi/2 \pm (4\,\eta_{0}^{2}+\delta)\,t$. Notice that $C \gg 1$ implies 
$(4 \eta_{0}^2+\delta)^2 \gg 16 \pi a \eta_{0}$, which can be solved for 
$\eta_{0}$, 
yielding for $a=0.05$, $\delta=-3$ and $\phi_{0}=\pi/2$,   
$\eta_{0} \ll 0.655$ or $\eta_{0} \gg 1.077$. 
This is confirmed by numerical solutions of the CC-equations which exhibit an oscillatory behavior of $\phi(t)$ 
for $\eta_{0} \in [0.65; 1.07]$, and a linear behavior (plus small oscillations) outside of this interval.

Finally we remark that in case II both $V_{s}$ in the stationary solutions and $\bar{V}$ 
in the oscillatory solutions differ from the value $2 K$ in case I. 
A transformation to a frame moving with $V \ne 2 K$ would not simplify 
the NLSE because terms with $\Psi_{X}$ would appear in Eq. (\ref{auto}). 
 
\section{Stability of oscillatory solutions} \label{seis}
 
Our simulations reveal that the driven undamped soliton is stable only for a part of 
the set of solutions obtained by the CC-theory. Naturally we would like to predict, by using the 
CC-theory, which solutions are unstable and to understand what causes the instability. 
(The latter point will be 
discussed in the next section).  

For the NLSE with a general local nonlinearity, but without driving and damping, the 
Vakhitov-Kolokolov stability criterion \cite{vk,w} states that solitons are stable if $dN/d\Lambda > 0$. 
Here 
$N$ is the norm and $\Lambda$ the so-called spectral parameter in stationary solutions of the 
form $u(x,t) = \Psi(x) \exp(i \Lambda t)$. 
However, our oscillatory solutions, 
Eqs.\ (\ref{eq1-ansatz})-(\ref{eq4-ansatz}) inserted 
into Eq.\ (\ref{aq}), have a more complicated time dependence than stationary solutions; 
therefore the criterion cannot 
be applied here. 

The same holds for the stability criteria of Barashenkov \cite{igor1,igor2} 
which were established 
for solitons travelling with constant velocity. But here we can get some motivation for how to 
proceed in our case: Barashenkov showed that dark solitons of the NLSE with 
generalized nonlinearity are stable if $d\tilde{P}/d\tilde{V}<0$ (here the tildes are used 
to distinguish from our $P$ and $V$)    
\cite{igor2}. This proof was carried over 
to (bright and dark) solitons of the undamped parametrically driven NLSE \cite{igor1}. Here 
the point $d\tilde{P}/d\tilde{V}=0$ separates a stable from an unstable branch of the curve 
$\tilde{P}(\tilde{V})$, but 
it depends on the type of the solution on which side the stable branch is. For the following it is 
important to note that this curve $\tilde{P}(\tilde{V})$ 
represents a family of solutions with different 
velocities, i.e. each solution is represented by one point on the curve. 

We make the conjecture that our oscillatory solutions are dynamically unstable, 
if our curve $P(V)$ has a branch with negative slope, i.e. 
\begin{equation} \label{curvepv}
\frac{dP}{dV} <0,    
\end{equation}
for a finite interval of $V$. This curve is obtained from its parameter representation $P(t)$, 
$V(t)$, where $P(t)= - 8 \eta \xi$ is the soliton momentum Eq.\ (\ref{ax}) and the soliton 
velocity $V(t)=\dot{\zeta}$ is obtained by the r.h.s. of Eq.\ (\ref{bc}). 
Each oscillatory solution is represented by its own curve $P(V)$. This curve has a finite length, because $P(t)$ 
and $V(t)$ are periodic and remain finite, ($\zeta(t)$ and $\phi(t)$, which contain terms linear 
in $t$, do not appear in $P(t)$, and in $V(t)$ they appear only via $\sin B$, see 
Eq.\ (\ref{bc})). Plotting the ``stability curve'' $P(V)$ we can inmediately see whether 
there is a branch with negative slope.  

For the parameter set $a=0.05$, $K=0.1$, $\beta=0$ and $\delta=-1$ and 
IC $\zeta_{0}=\xi_{0}=0$ and $\phi_{0}=\pi/2$, we find 
a small ``stability interval'' $0.48 \le \eta_{0} \le 0.52$, i.e. an interval 
of initial conditions for which the solutions are stable. As expected, this interval is 
situated around the value $\eta_{s}=0.501874$ from the stable stationary solution (\ref{eq-eta}). 
There is another stable regime for 
$\eta_{0} \ge 0.76$. When we go far away from the IC for 
the stationary solution by choosing $\phi_{0}=0$, instead of $\phi_{0}=\pi/2$, the stability interval around $\eta_{s}$ 
vanishes. The upper stability regime exists now for    $\eta_{0} \ge 0.69$. 

When $|\delta|$ is increased, e.g. by choosing $\delta=-3$, the stability interval 
around $\eta_{s}=0.866239$ is much 
larger than in the case $\delta=-1$; for $\phi_{0}=\pi/2$ it is $0.7 \le \eta_{0} \le 1.03$. The upper stability region   
is above $\eta_{0}=1.08$. Considering again $\phi_{0}=0$, the stability interval around 
$\eta_{s}$ only shrinks 
to $0.76 \le \eta_{0} \le 0.97$, but does not vanish because it was much larger than in the case $\delta=-1$. The upper stable region is above $\eta_{0}=1.02$. Thus the 
conclusion is that 
{\it an increase of 
$|\delta|$ widens the regions with stable soliton solutions}. All the stability regions given above are confirmed by simulations for the perturbed NLSE, with an error of less than $1\, \%$. 

\begin{figure}[ht]
\begin{center}
\begin{tabular}{cc}\ & \\
\includegraphics[width=7.5cm]{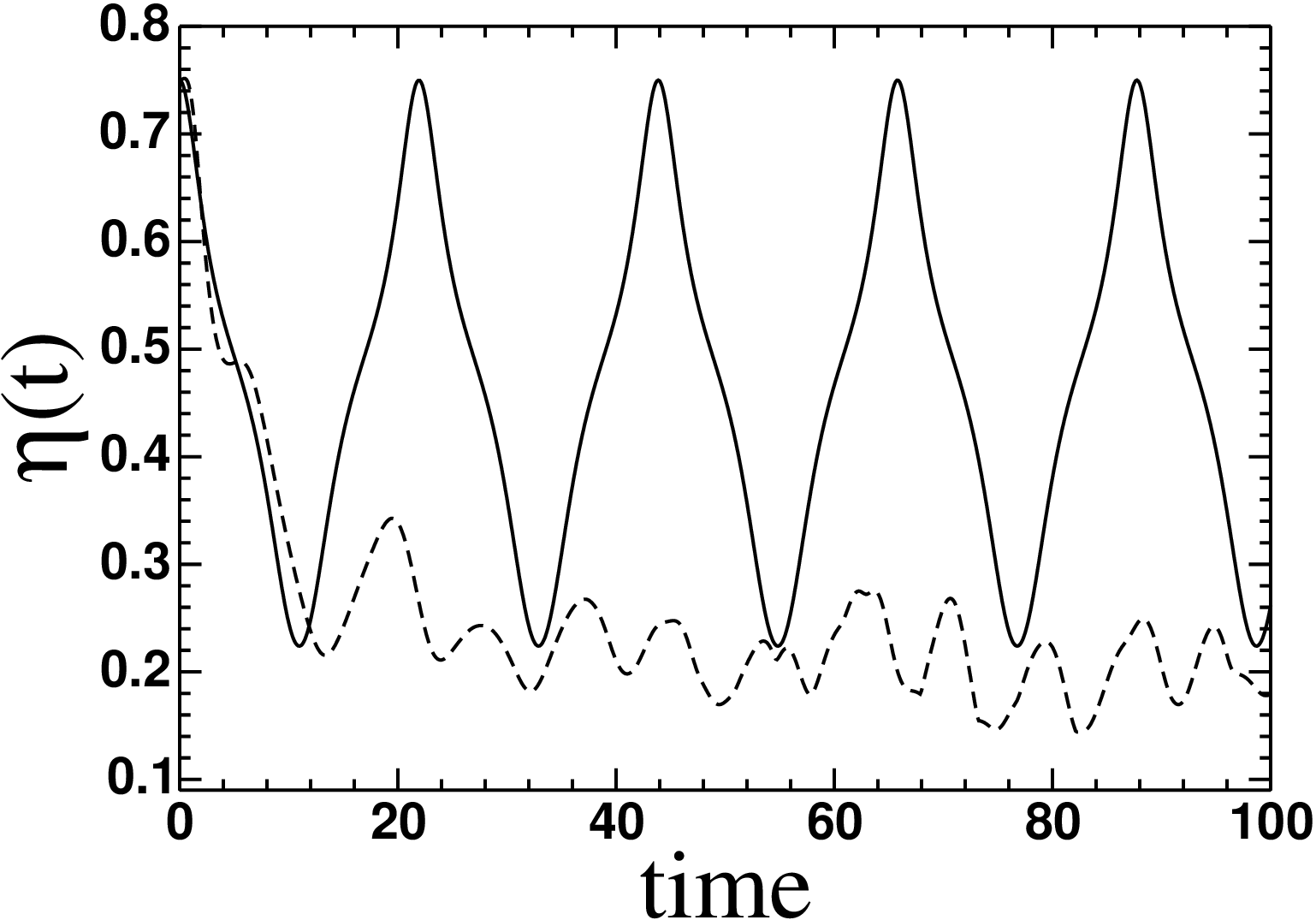} & 
\quad \includegraphics[width=7.5cm]{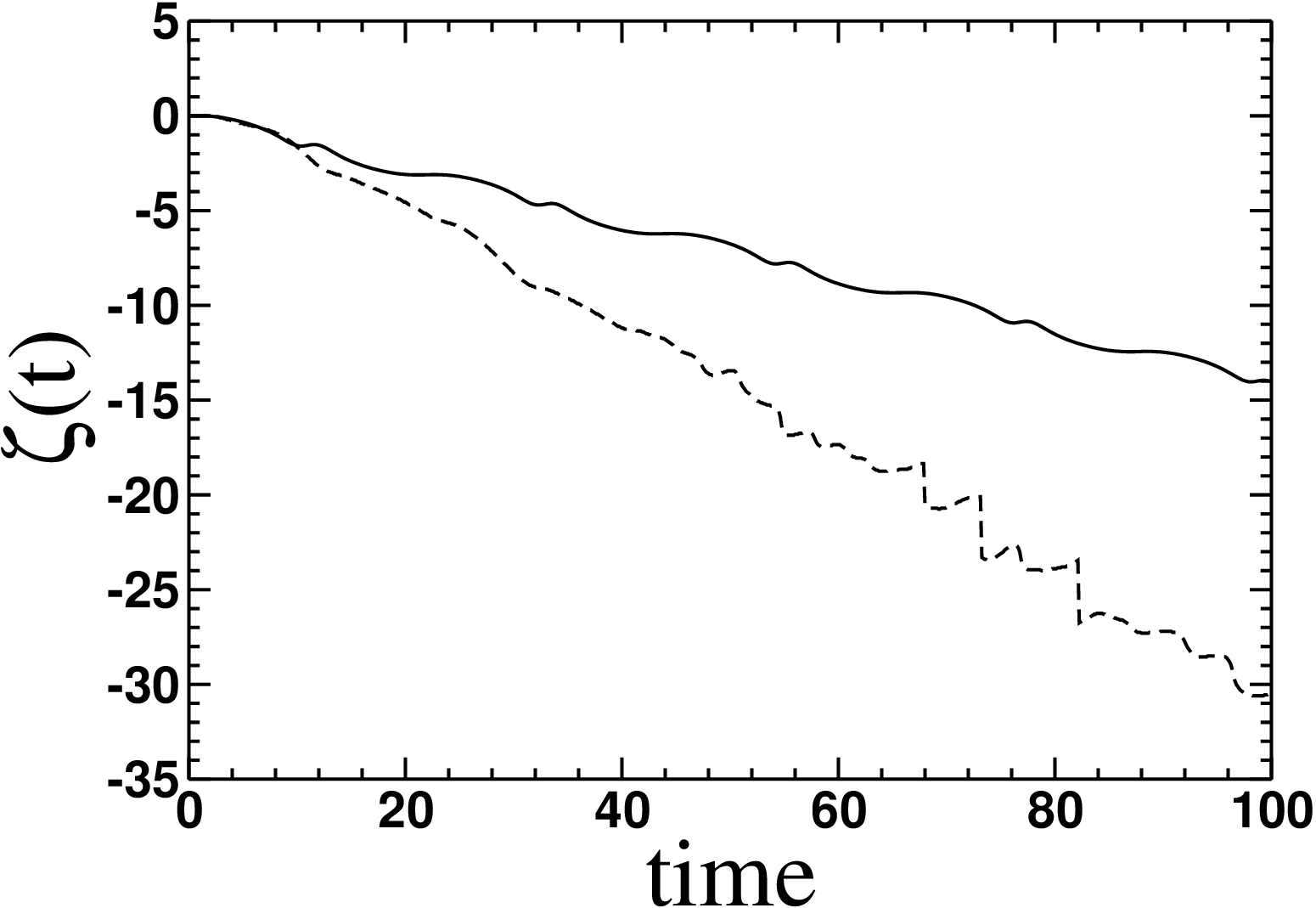} \\
\\
\\
\includegraphics[width=7.5cm]{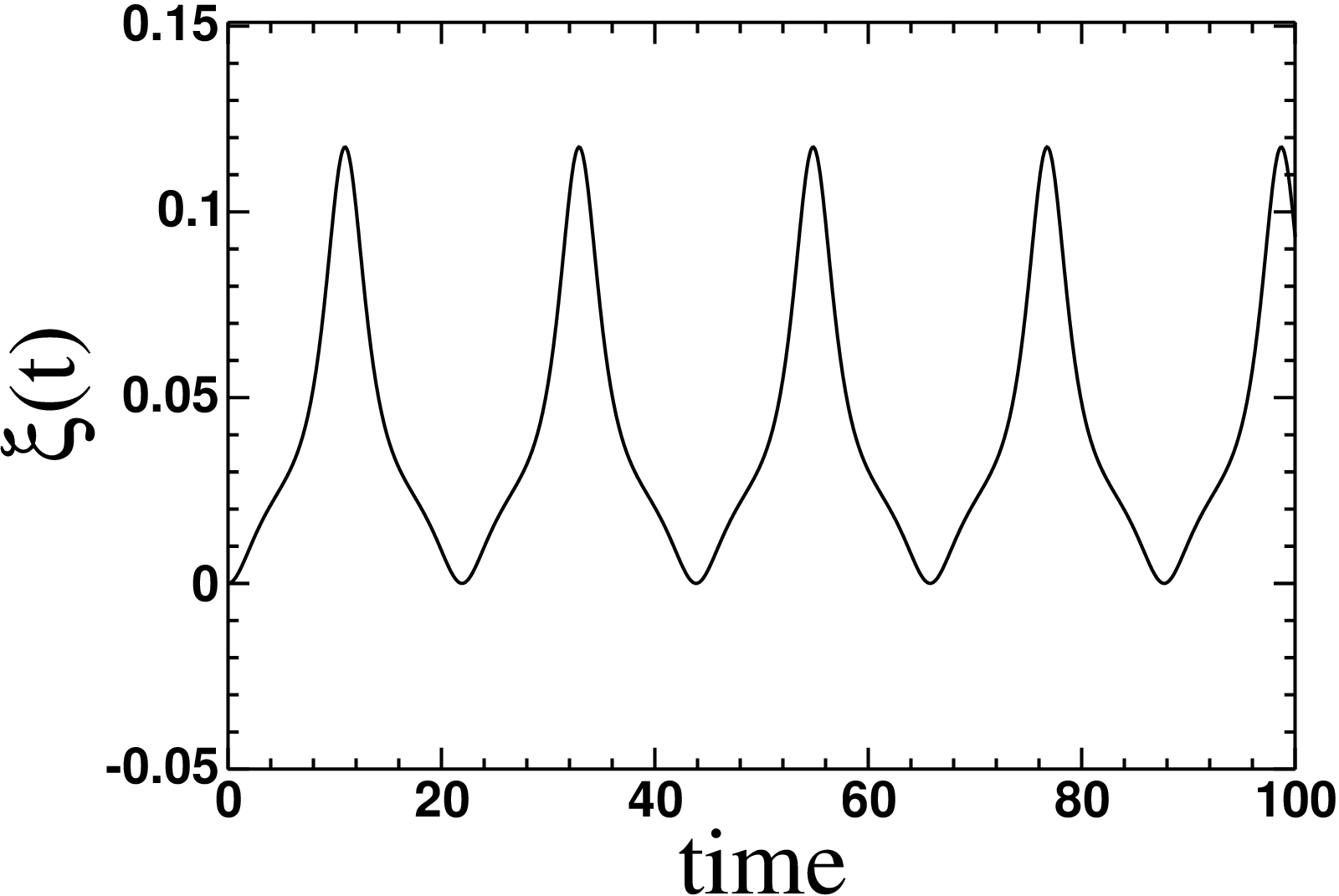} & 
\quad \includegraphics[width=7.0cm]{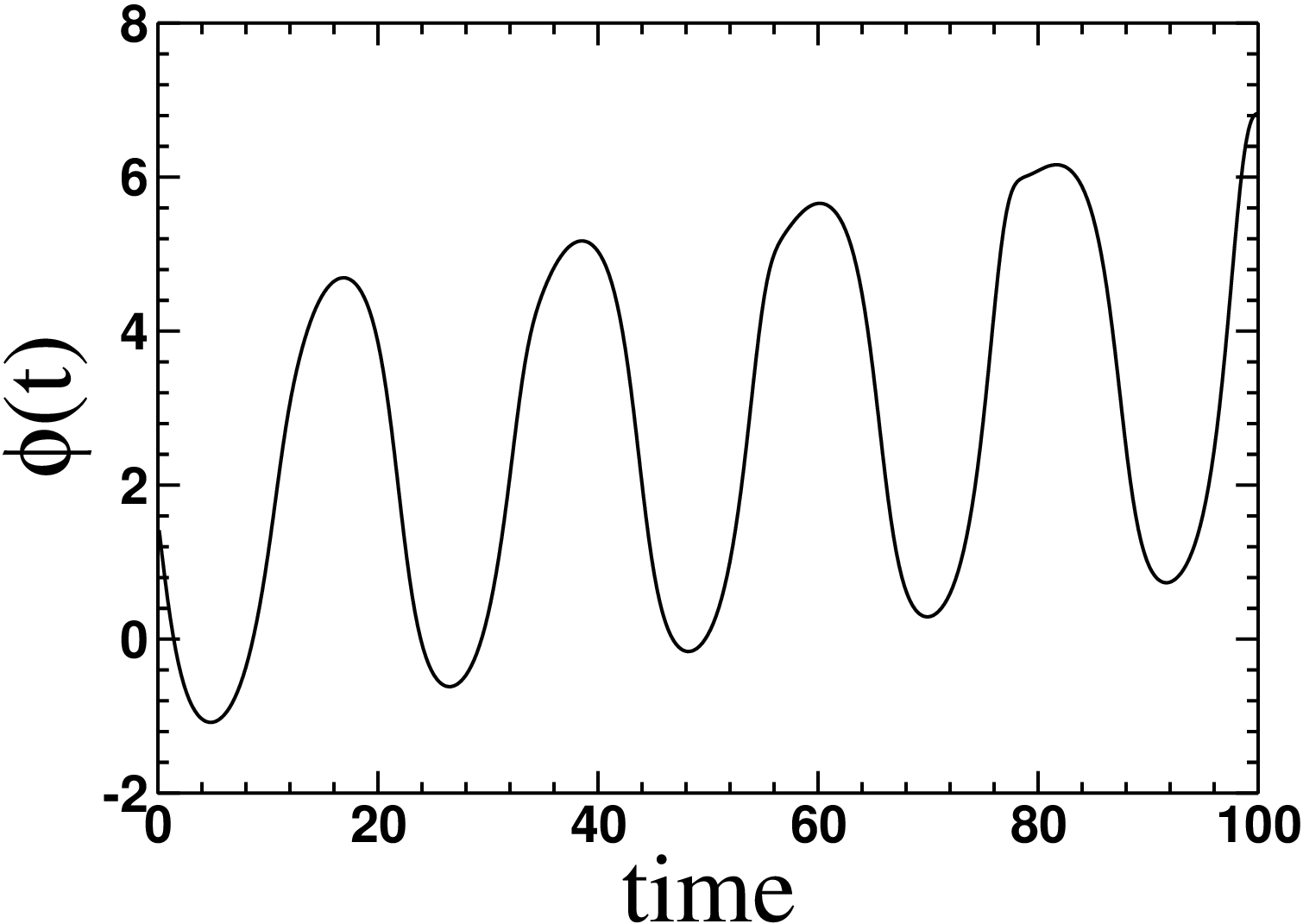} \\
\end{tabular}
\end{center}
\caption{ 
 The evolution of $\eta$, $\zeta$, $\xi$ and $\phi$  
obtained from a  simulation of the NLSE (dashed lines) 
and from the numerical solutions of the CC equations (solid lines). 
Parameters: $K=0.1$, $a=0.05$, $\delta=-1$, $\beta=0$, with IC $\xi_{0}=0$, $\zeta_{0}=0$, 
$\phi_{0}=\pi/2$ and $\eta_{0}=0.75$.  
} 
\label{fig3}
\end{figure}

\begin{figure}[ht]
\begin{center}
\begin{tabular}{cc}\ & \\
\includegraphics[width=7.5cm]{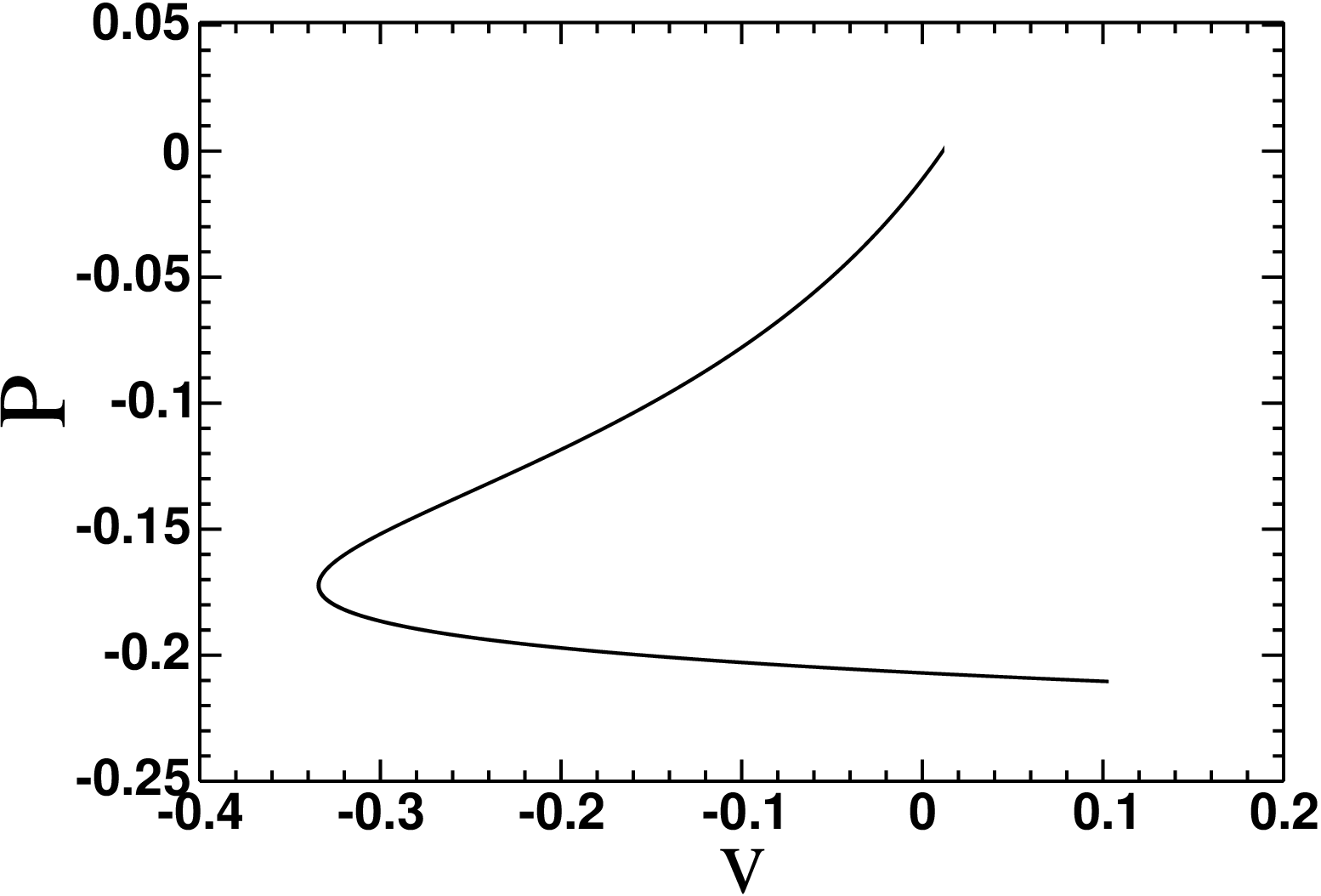} & 
\quad \includegraphics[width=7.5cm]{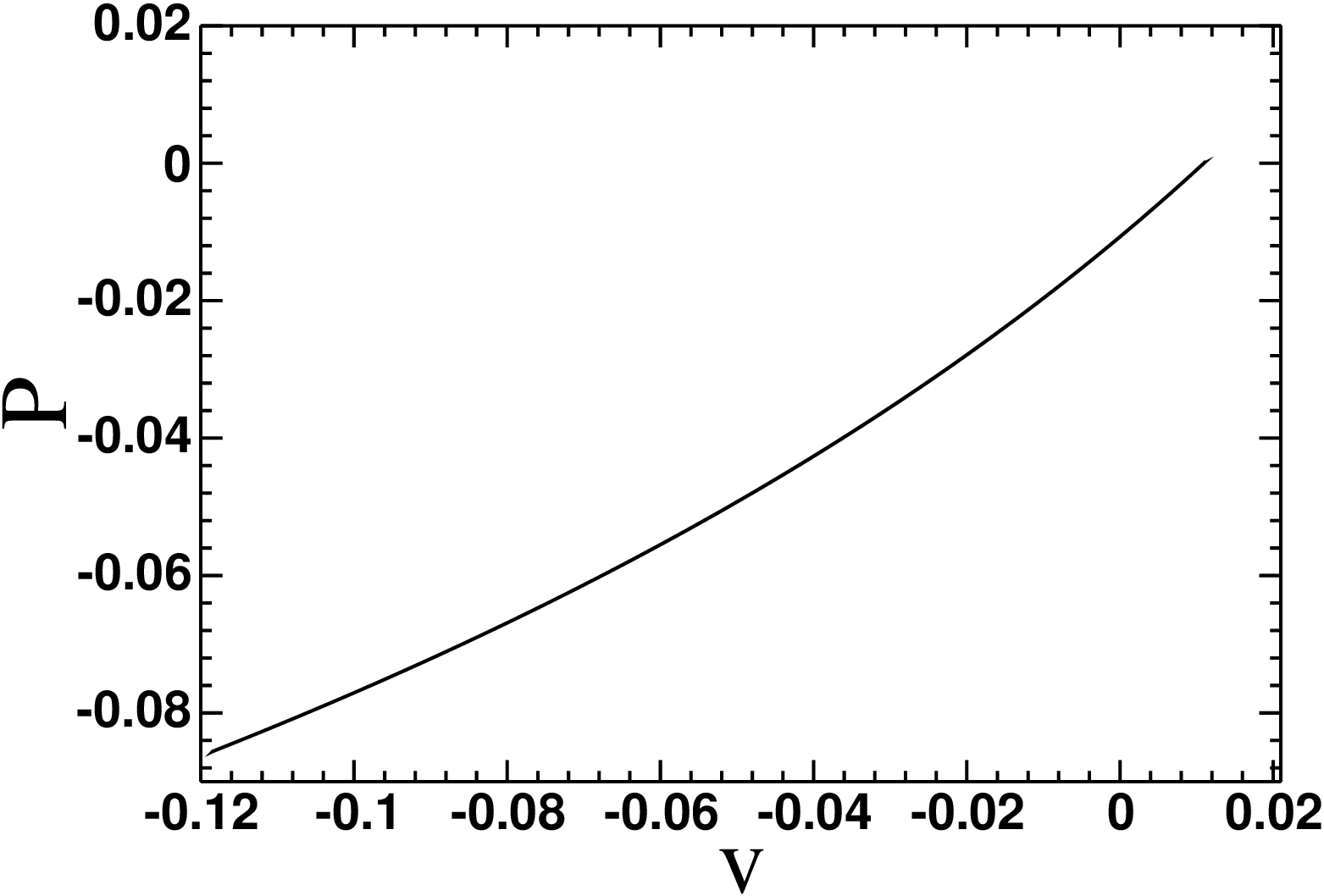} 
\end{tabular}
\end{center}
\caption{``Stability curve'' $P$ versus $V$. Left panel:  $\eta_{0}=0.75$. Right panel $\eta_{0}=0.76$.  
Other parameters as in Fig.\ \ref{fig3}. 
} 
\label{fig4}
\end{figure}

\begin{figure}[ht]
\begin{center}
\begin{tabular}{cc}\ & \\
\includegraphics[width=7.5cm]{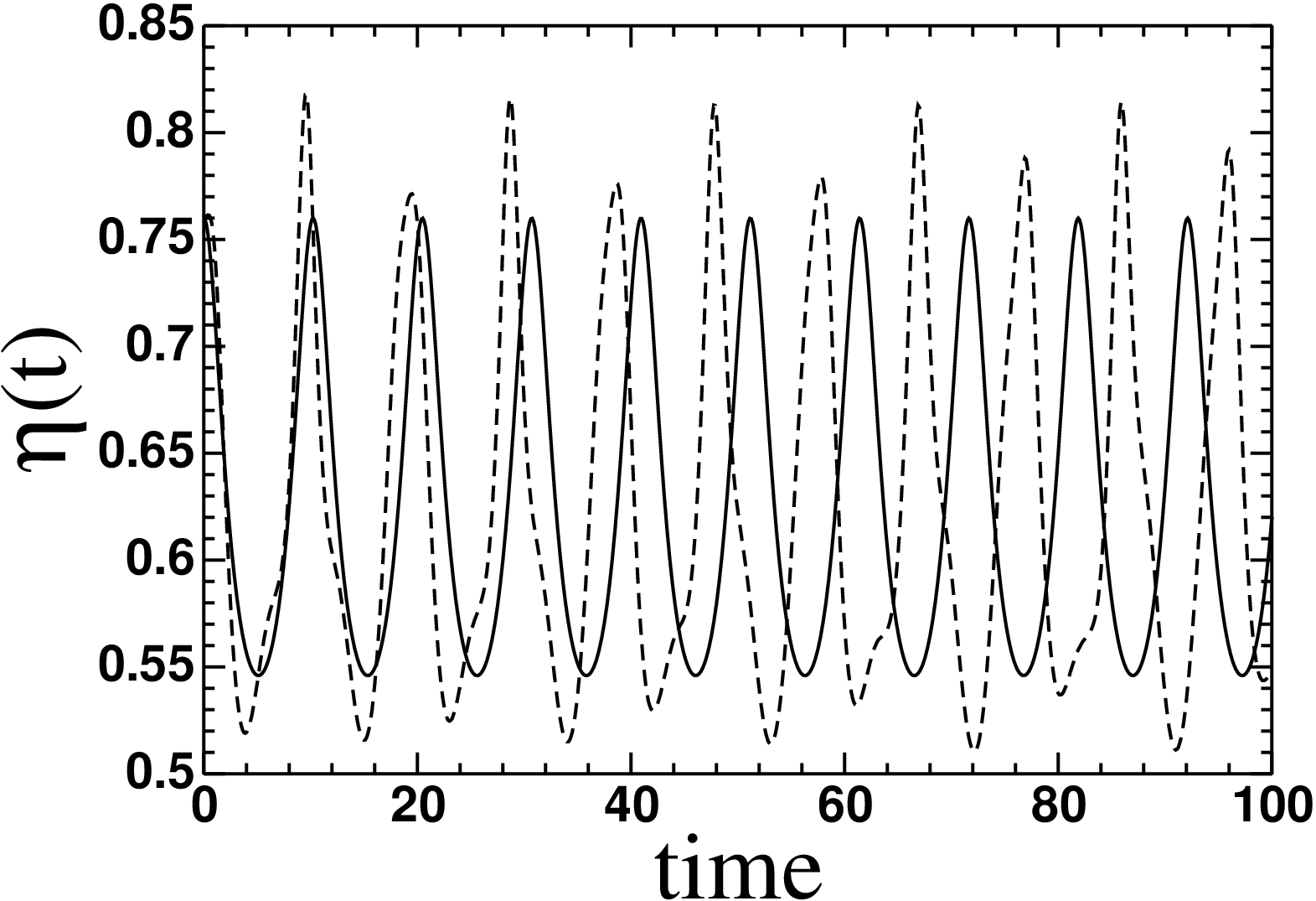} & 
\quad \includegraphics[width=7.5cm]{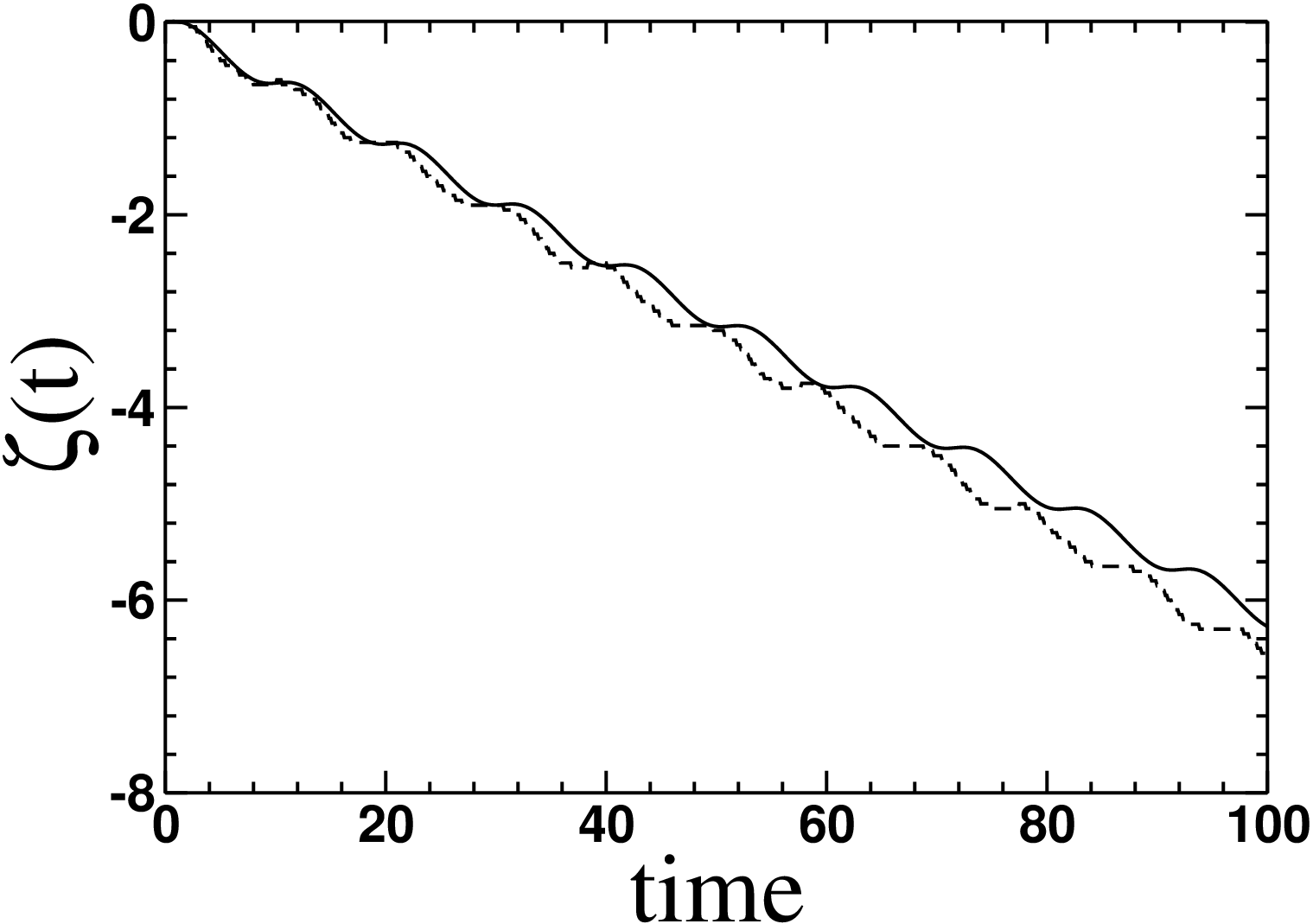} \\
\\
\\
\includegraphics[width=7.5cm]{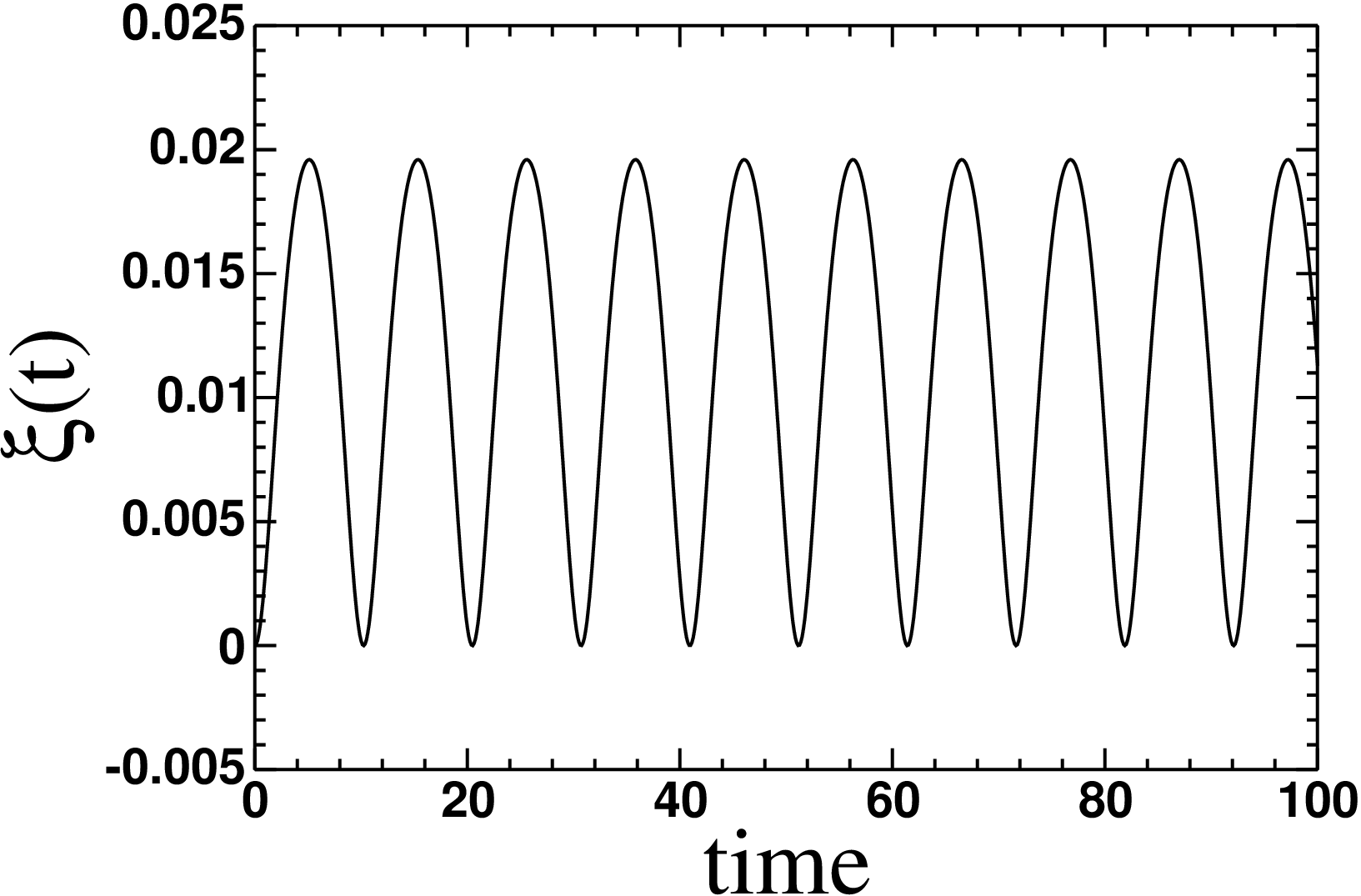} & 
\quad \includegraphics[width=7.0cm]{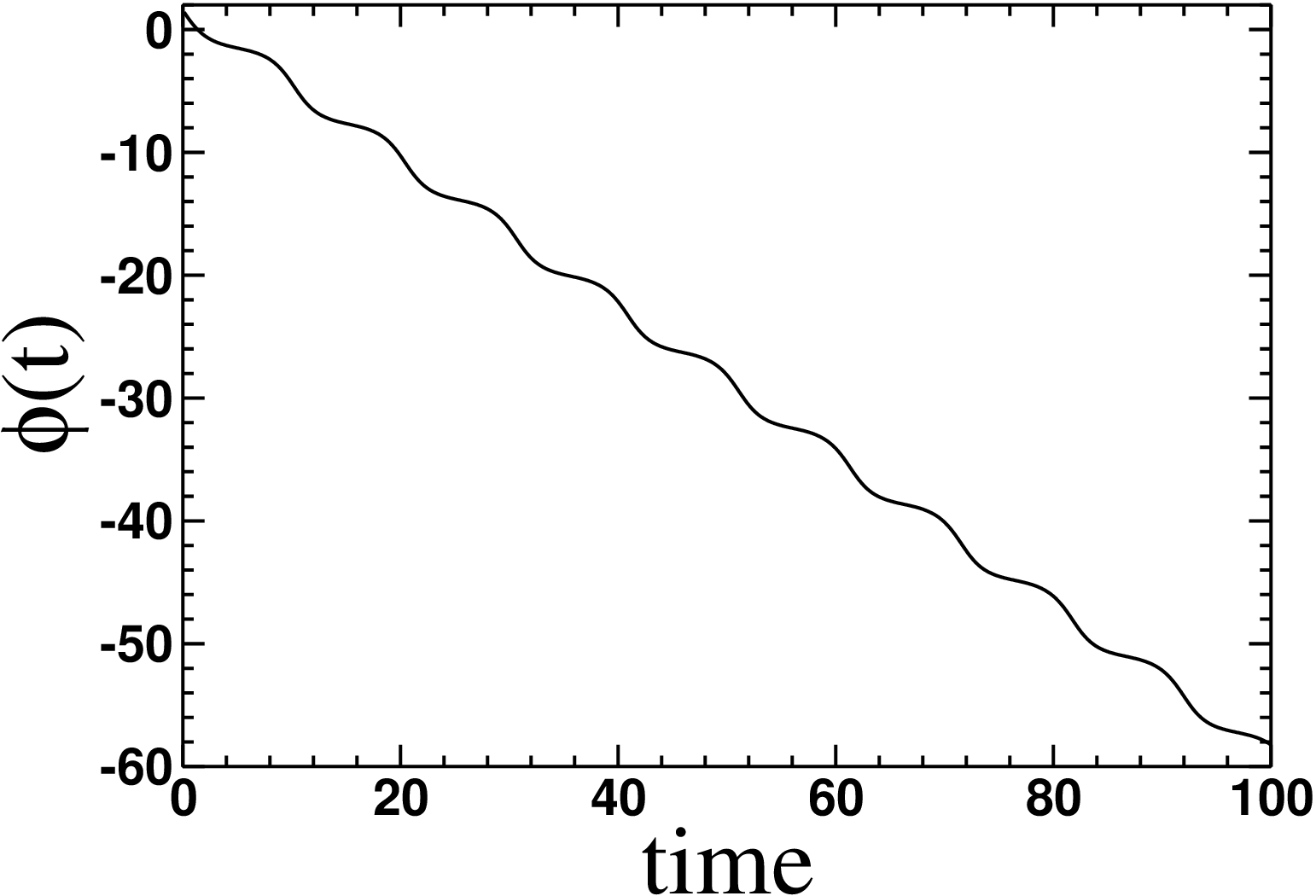} \\
\end{tabular}
\end{center}
\caption{ 
 The evolution of $\eta$, $\zeta$, $\xi$ and $\phi$  
obtained from a simulation of the NLSE (dashed lines) 
and from the numerical solutions of the CC equations (solid lines).  
Parameters: $K=0.1$, $a=0.05$, $\delta=-1$, $\beta=0$, with IC $\xi_{0}=0$, $\zeta_{0}=0$, 
$\phi_{0}=\pi/2$ and $\eta_{0}=0.76$. The simulations were carried forward to a final time 
$t_{f}=1000$.   
} 
\label{fig5}
\end{figure}

At some of the boundaries of the above stability intervals there is a drastic change in 
the shape of the solutions of the CC-equations and the stability curve. E.g., if we choose 
$\eta_{0}=0.75$ (Fig.\ \ref{fig3}), which is just below the stability regime $\eta_{0} \ge 0.76$ (see above), 
we obtain very anharmonic oscillations in all CCs 
and the stability curve has a long branch with negative slope 
(Fig.\ \ref{fig4}). The simulations indeed show 
that the soliton becomes unstable very quickly and vanishes (Fig.\ \ref{fig3}). However, a slight change 
of $\eta_{0}$ to the value $0.76$ produces a stability curve that has only one branch with a positive slope  (Fig.\ \ref{fig4}). 
Here the soliton is indeed stable in the simulations (Fig.\ \ref{fig5}). 

\begin{figure}[ht]
\begin{center}
\begin{tabular}{c}\  \\
\includegraphics[width=7.5cm]{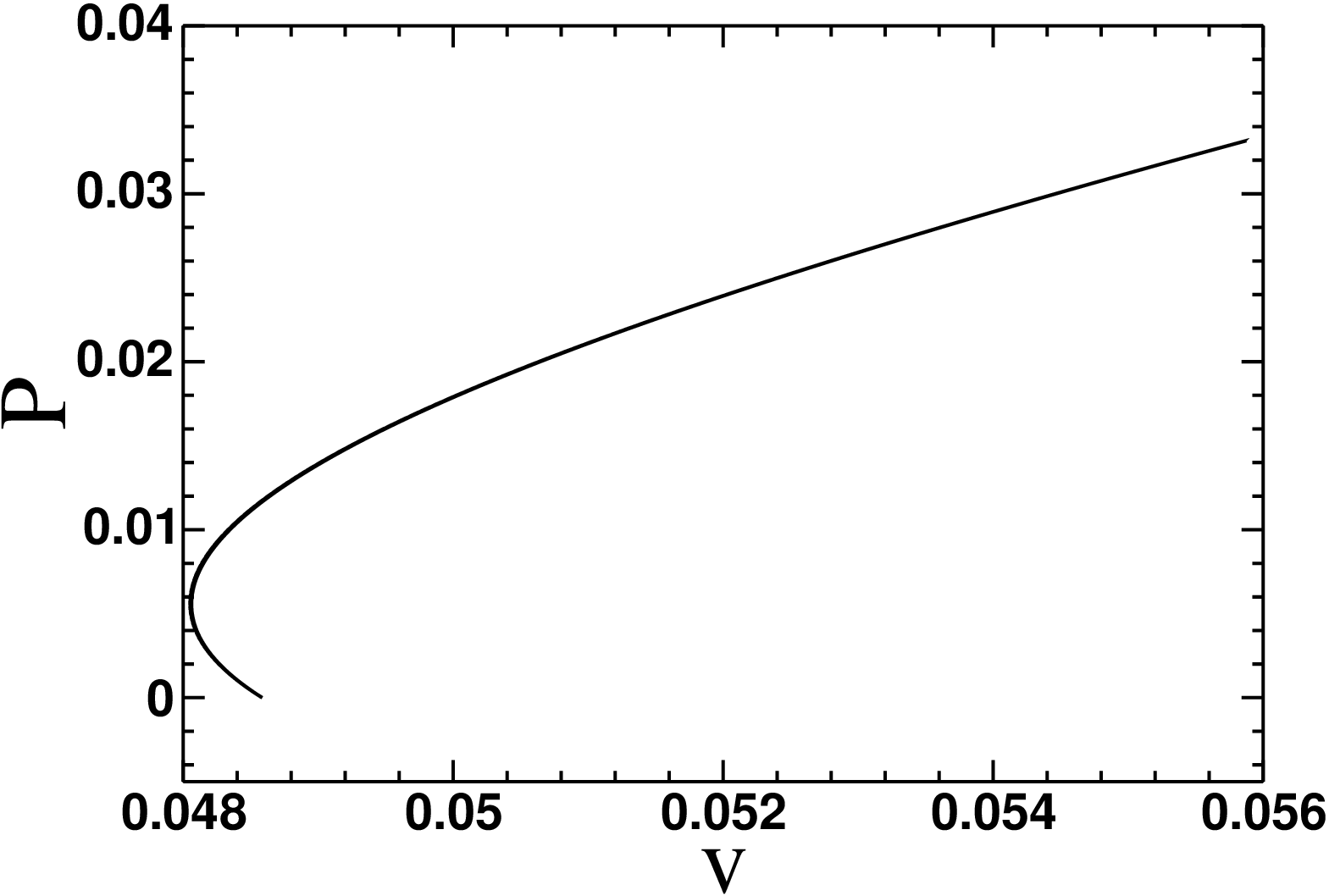} 
\end{tabular}
\end{center}
\caption{ 
 ``Stability curve'' close to the boundary of an unstable regime.   
Parameters: $K=0.1$, $a=0.05$, $\delta=-1$, $\beta=0$, with IC $\xi_{0}=0$, $\zeta_{0}=0$, 
$\phi_{0}=\pi/2$ and $\eta_{0}=0.46$. 
} 
\label{fig6}
\end{figure}

\begin{figure}[ht]
\begin{center}
\begin{tabular}{cc}\ & \\
\includegraphics[width=7.5cm]{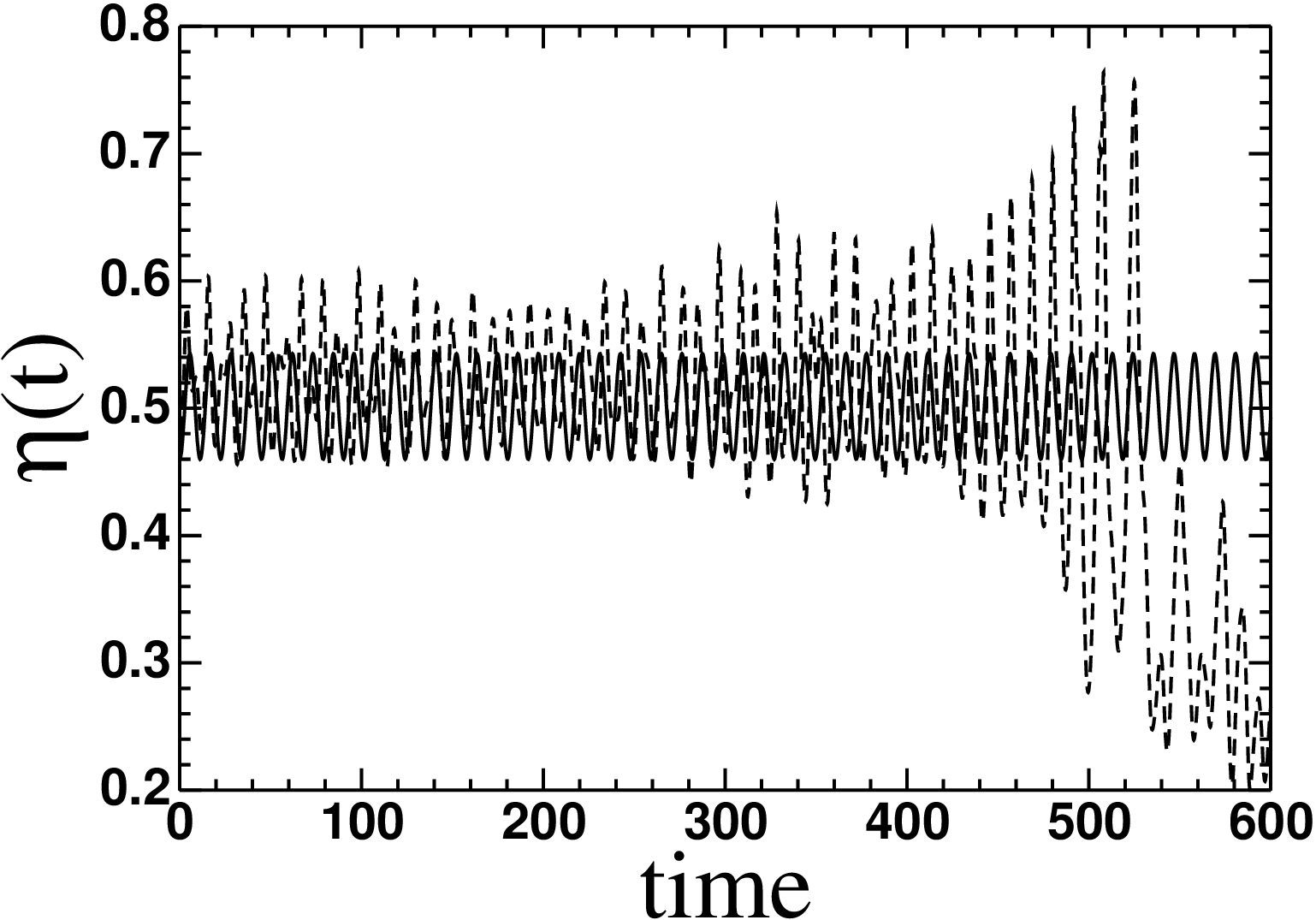} & 
\quad \includegraphics[width=7.5cm]{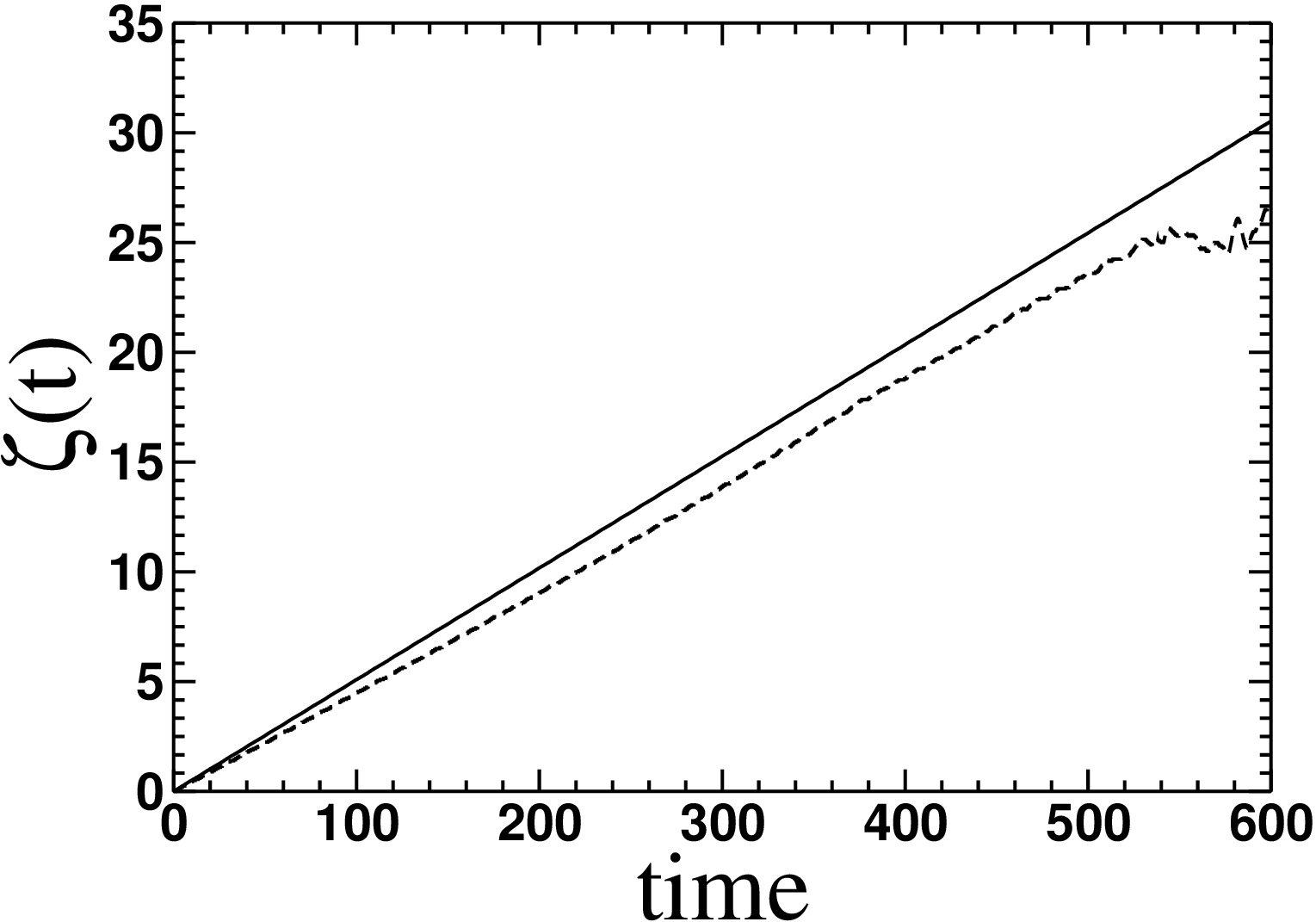} \\
\\
\\
\includegraphics[width=7.5cm]{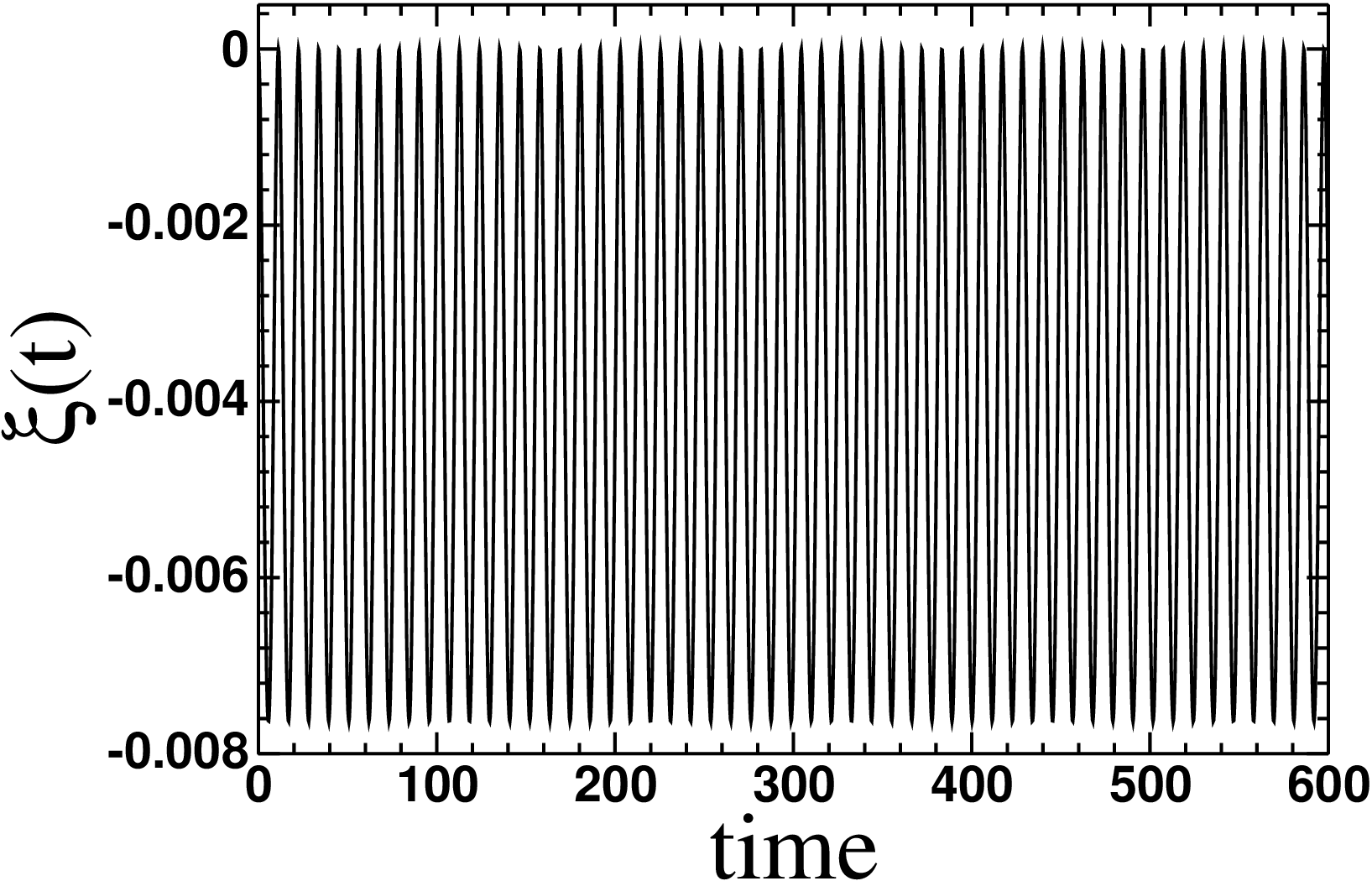} & 
\quad \includegraphics[width=7.0cm]{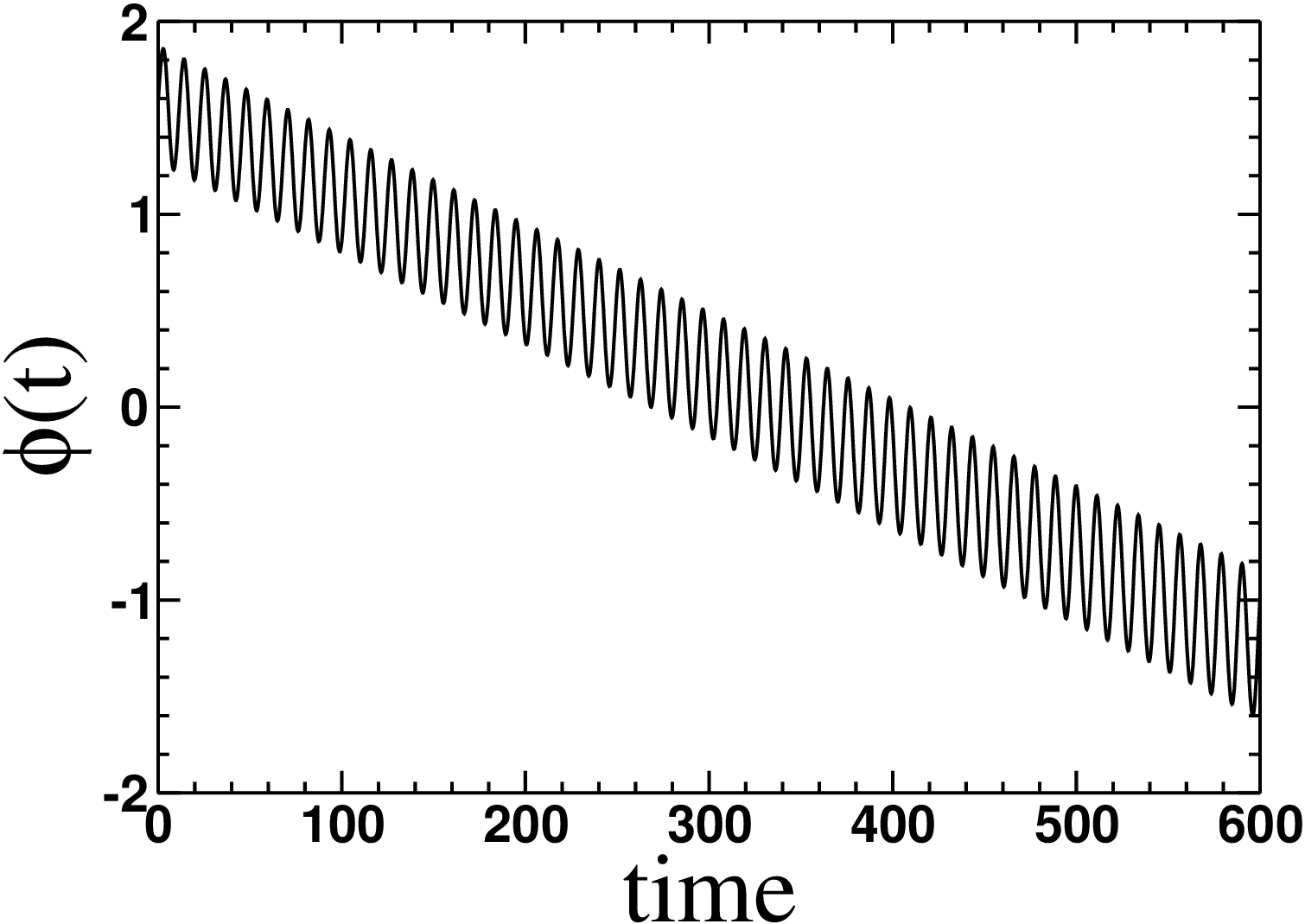} \\
\end{tabular}
\end{center}
\caption{ 
 The evolution of $\eta$, $\zeta$, $\xi$ and $\phi$  
obtained from a simulation of the NLSE (dashed lines) 
and from the numerical solutions of the CC equations (solid lines). Parameters as in Fig.\ \ref{fig6}.  
} 
\label{fig7}
\end{figure}

At the boundaries of other stability intervals the change of the shape of the solutions is less dramatic. E.g., for the case 
$\delta=-1$ and $\eta_{0}=0.46$ the oscillations in the CCs are small and harmonic. The stability curve has only a very short branch with negative slope which is visited in the time evolution only for very short time 
intervals (Fig.\ \ref{fig6}). Here the soliton is indeed stable for a relatively long time (Fig.\ \ref{fig7}). This soliton 
lifetime is increasingly reduced when $\eta_{0}$ is reduced by which the negative slope branch in 
$P(V)$ becomes longer. {\it Thus this curve predicts not only 
whether the soliton is stable, but also gives an estimate for its lifetime when it is unstable.}

\section{Soliton and phonon dispersion curves} \label{siete} 

For the soliton 
dispersion curve $E^{sol}(P)$ we need the canonical momentum $P$ of the soliton. 
$P$ must fulfill the Hamilton equation
\begin{equation} \label{eq-H}
\dot{\zeta}= \frac{\partial H}{\partial P},
\end{equation}
where $\zeta$ is the soliton position. We start from the CC-Lagrangian Eq.\ 
(\ref{bh}) and obtain
\begin{equation} \label{eq-lphi}
\frac{\partial L}{\partial \dot{\phi}} = 4 \eta = N,
\end{equation}
as the angular momentum of the internal oscillation of the soliton, and 
\begin{equation} \label{eq-lxi}
\frac{\partial L}{\partial \dot{\xi}} = 8 \eta \zeta = 2 N \zeta :=X,
\end{equation}
as the variable which is canonically conjugate to $\xi$. Using a Legendre transformation 
we obtain the Hamilton function 
\begin{eqnarray} \label{eq-hamilton}
H&=& X \dot{\xi} + N \dot{\phi} - L, \\
 \label{eq-hamilton1}
H(X,\xi;N,\phi)&=& 4 N \xi^{2} - \frac{1}{12} N^{3} - \delta N + 2 \pi a \,  
\mathrm{sech} A \cdot \sin B,
\end{eqnarray}
with 
\begin{equation} \label{eq-AyB}
A=2 \pi\,\frac{K+2 \xi}{2 N}, \qquad B=\phi+ \frac{K+2 \xi}{2 N}\,X.
\end{equation}
We now make an ansatz for a canonical transformation to the following new set of variables: 
\begin{eqnarray} \label{eq-canonical}
P &=& - 2 N \xi, \quad \zeta=\frac{1}{2 \, N} \, X \\
\tilde{N}&=&N,    \quad \tilde{\phi}=\phi+g(N,\xi,X).
\end{eqnarray}
This means that $P$ is identical to the kinetic momentum in Eq.\ (\ref{ax}), $\tilde{N}$ is chosen to be equal to $N$ 
and $g$ still has to be determined such that the four fundamental Poisson brackets are fulfilled. This yields 
$g=\xi X/N$. The new Hamiltonian reads
\begin{equation} \label{eq-hamilton2}
H(P,\zeta;N,\tilde{\phi})= \frac{1}{N} P^{2} - \frac{1}{12} N^{3} - \delta N + 2 \pi a \, 
\mathrm{sech} A \cdot \sin B,
\end{equation}
with 
\begin{equation} \label{eq-AyB1}
A=\frac{\pi}{N} \, \left(K-\frac{P}{N}\right), \qquad B=\tilde{\phi}+ K \zeta.
\end{equation}
The first two terms in Eq.\ (\ref{eq-hamilton2}) agree with literature results on the unperturbed NLSE \cite{fadd}. 
We have checked that the four Hamiltonian Eqs.\ which result from Eq.\  (\ref{eq-hamilton2}) are indeed equivalent 
to the four CC-equations in section \ref{tres} for the case without damping.
The above results hold for the force $f(x,t)$ in Eq.\ (\ref{ac}) with arbitrary $K(t)$. 
In the following we return to the case of constant $K$ in order to calculate analytically 
the stability and dispersion curves wherever it is possible.  

Let us consider the stability intervals around the stationary solutions as given in Section 
\ref{seis}. Except for the close vicinity to the boundaries of the stability intervals, 
the oscillations in the CCs are nearly harmonic and 
can be well approximated by the Eqs.\ (\ref{eq1-ansatz}) to (\ref{eq4-ansatz}). Inserting into $P=-8\,\eta\,\xi$, 
neglecting the 2nd harmonic, and using $V=\dot{\zeta}$, one can easily see that $P(V)$ is a straight line with slope 
\begin{equation} \label{eqdpdv}
\frac{dP}{dV}=\frac{8\,\eta_{0}a_{\xi}}{\Omega\,a_{\zeta}} >0,
\end{equation}
because for $\eta_{0}<\eta_{s}$ both $a_{\xi}$ and $a_{\zeta}$ are positive and for 
$\eta_{0}>\eta_{s}$ both are negative. The latter also holds for the upper stability 
intervals $\eta_{0} \ge 0.76$ 
for $\delta=-1$ and $\eta_{0} \ge 1.08$ for $\delta=-3$. Near or at the boundaries of the 
stability intervals, the oscillations in the CCs are very anharmonic and therefore 
the calculation leading to Eq.\ (\ref{eqdpdv}) is not possible. $P(V)$ is a curved line 
which can be calculated using the numerical solution of the CC-equations.   

We now turn to the soliton dispersion curve $E^{sol}(P)$. Both $P=-8\,\eta\,\xi$ and $E^{sol}$ in Eq.\ (\ref{aw}) 
consist of powers of $\eta$ and $\xi$. 
For simplicity we write Eqs.\ (\ref{eq2-ansatz}) and (\ref{eq3-ansatz}) as 
$\eta = \bar{\eta} - a_{\eta}\,\cos(\Omega \,t)$ and 
$\xi = \bar{\xi} + a_{\xi}\,\cos(\Omega \,t)$, where 
$\bar{\xi}$ is negligible. 
We concentrate on the leading terms in 
$\cos(\Omega\,t)$ and distinguish two cases: 
In case I, both, $E^{sol}(t)$ and $P(t)$ have a leading term with 
$\cos(\Omega\,t)$; in this case the dispersion curve $E^{sol}(P)$ is linear in the first approximation. In case II, the 
$1^{st}$-order terms in $E^{sol}(t)$ cancel if $\bar{\eta}=\frac{1}{2}\,\sqrt{-\delta}$, but cannot cancel in $P(t)$. 
In the next order 
\begin{equation}\label{eq-d1}
E^{sol}(t)=E_{max}-\Delta E \,\cos^{2}(\Omega\,t),
\end{equation}
with $\Delta\,E=16\,\bar{\eta}(a_{\eta}^{2}/3-a_{\xi}^{2}) > 0$ because 
$|a_{\xi}| \ll |a_{\eta}|$. Inserting 
$P=\bar{P} + 8 \bar{\eta} a_{\xi} \cos(\Omega\,t)$ yields a parabolic dispersion curve
\begin{equation}\label{eq-d2}
E^{sol}(t)=E_{max}-\Delta E \,\left(\frac{P-\bar{P}}{8\,\bar{\eta}\,a_{\xi}}\right)^{2}.
\end{equation}
This turns out to be a surprisingly good approximation when comparing with the dispersion curve obtained by 
using numerical solutions for $\eta(t)$ and $\xi(t)$, even when the cancellation of the linear terms in $E^{sol}(t)$ 
is not exact. The condition $\bar{\eta}=\sqrt{-\delta}/2$ is approximately fulfilled for the regions with strong oscillatory 
terms in $\phi(t)$; see end of Section \ref{cinco}.  

When solitons become unstable they radiate phonons (i.e., linear excitations). 
Therefore we consider the perturbed linearized NLSE without damping
\begin{equation} \label{nlse1}
iu_t + u_{xx} + \delta u = a\,  e ^{i K x },
\end{equation} 
which is solved by 
\begin{equation} \label{nlse2}
u(x,t) = c\,  e ^{i (k x -\omega_{k} t)} +  b\,  e ^{i (K x -\omega_{K} t)} - \frac{a}{\omega_{K}}\,  e ^{i K x }
\end{equation}
with $\omega_{k}=k^2-\delta$, $\omega_{K}=K^2-\delta$. 
The first term in Eq.\ (\ref{nlse2}) represents the phonons 
of the unperturbed equation with the free amplitude $c$ and the dispersion curve $\omega_{k}$. 

The second term in Eq.\ (\ref{nlse2}) with the free amplitude $b$ represents a single phonon mode whose wave number and frequency are given by the parameter $K$ in the force 
$f(x)=a \, e ^{i K x }$. Such phonons are typically radiated at the beginning of a simulation 
as the initial soliton profile adapts to the system. These phonons can be observed best when they interact with the soliton after having been reflected by a boundary of the system. 

Finally, the last term in Eq.\ (\ref{nlse2}) represents a static background with a fixed amplitude $a/\omega_{K}$. 
This was already discussed below Eq.\ (\ref{backg}).
 
\section{Conclusions}

We have considered the dynamics of NLS-solitons in one spatial dimension under the influence of non-parametric spatio-temporal forces of the form $f(x,t)=a\, \exp[i\,K(t)\,x]$, plus a damping term and a linear term $\delta\,u(x,t)$ which stabilizes the driven soliton. We have developed a CC-theory which yields a set of ODEs for the four CCs (position 
$\zeta$, velocity $\xi$, 
amplitude $\eta$, and phase $\phi$). 

These coupled ODEs have been solved analytically and numerically for the case of a constant, spatially periodic force 
$f(x)=a\, \exp[i\,K\,x]$. The soliton position exhibits oscillations around a mean trajectory 
$\bar{\zeta}=\bar{V}\,t$; this means that the soliton performs, on the average, 
a unidirectional motion although the spatial average of the force vanishes. 
The amplitude of the oscillations is much smaller than the spatial period $L=2\pi/|K|$ of the inhomogeneity $f(x)$. The other three CCs 
also exhibit oscillations with the same frequency as $\zeta(t)$. 

In the case of damping, the above oscillations are damped and the solution approaches a steady-state solution with 
constant velocity, if the IC are close enough to those of the steady-state solution and if the damping is not too large. Otherwise the 
soliton vanishes, i.e. its amplitude and energy go to zero while its width goes to infinity. 

In the case without damping all the above oscillations persist. These periodic solutions exist because the total energy 
of the {\it perturbed} system is a conserved quantity, even for {\it arbitrary} inhomogeneity $f(x)$, and independent 
of the CC-ansatz.   

However, a comparison with simulation results for the perturbed NLSE reveals that only part of the above 
oscillatory solutions are stable. Our CC-theory predicts the unstable regions in the IC and the parameter $\delta$ with high 
accuracy, 
by using our conjecture that the soliton becomes unstable if 
 the slope of the 
curve $P(V)$ becomes negative somewhere: here $P(t)$ and $V(t)$ 
are the soliton momentum and velocity, respectively. 
It turns out that the stability intervals become broader when the parameter $\delta$ is chosen more  
negative. Moreover, we have found that the curve $P(V)$ also yields a good estimate for the soliton lifetime: The 
soliton lives longer, the shorter the negative-slope branch is, as compared to the length of the positive-slope branch. 

Other cases of the force $f(x,t)=\exp[i\,K(t)\,x]$ will be considered in a second paper: specifically single and biharmonic 
$K(t)$, with and without damping.

\section{Acknowledgments}
We thank Yuri Gaididei (Kiev) and Igor Barashenkov (Cape Town) 
for very useful discussions  
on this work. F.G.M. acknowledges the hospitality of the University of Sevilla 
and of the Theoretical 
Division and Center for Nonlinear Studies at Los Alamos Laboratory. Work at Los Alamos is 
supported by the USDOE.  
F.G.M. acknowledges financial support from IMUS and from University of Seville (Plan Propio).
N.R.Q. acknowledges financial support 
by the Ministerio de Educaci\'on y Ciencia (MEC, Spain) 
through FIS2008-02380/FIS, and by the Junta de Andaluc\'{\i}a
under the projects FQM207, FQM-00481 and  P06-FQM-01735.

\end{document}